%% file: paper_arxiv.tex
\documentclass[12pt]{article}

\include{_preamble}

\AtEndDocument{\refstepcounter{theorem}\label{finalthm}}
\AtEndDocument{\refstepcounter{equation}\label{finaleq}}

\title{\titlepaper}

\begin{document}
\maketitle

\begin{abstract}
  Most causal inference methods consider counterfactual variables
  under interventions that set the exposure to a fixed value. With
  continuous or multi-valued treatments or exposures, such
  counterfactuals may be of little practical interest because no
  feasible intervention can be implemented that would bring them
  about. Longitudinal modified treatment policies (LMTPs) are a
  recently developed non-parametric alternative that yield effects of
  immediate practical relevance with an interpretation in terms of
  meaningful interventions such as reducing or increasing the exposure
  by a given amount. LMTPs also have the advantage that they can be
  designed to satisfy the positivity assumption required for causal
  inference. We present a novel sequential regression formula that
  identifies the LMTP causal effect, study properties of the LMTP
  statistical estimand such as the efficient influence function and
  the efficiency bound, and propose four different estimators. Two of
  our estimators are efficient, and one is sequentially doubly robust
  in the sense that it is consistent if, for each time point, either
  an outcome regression or a treatment mechanism is consistently
  estimated. We perform numerical studies of the estimators, and
  present the results of our motivating study on hypoxemia and
  mortality in intubated Intensive Care Unit (ICU) patients. Software
  implementing our methods is provided in the form of the open source
  \texttt{R} package \texttt{lmtp} freely available on GitHub
  (\url{https://github.com/nt-williams/lmtp}) and CRAN.

\end{abstract}

\section{Introduction}






Most modern causal inference methods use a potential outcome framework
to define causal effects as contrasts between the outcome distribution
under different hypothetical counterfactual worlds. The defining
characteristic of these counterfactual worlds is a series of
interventions on the cause being evaluated. For example, causal
effects for binary variables can be defined as the difference in mean
outcome comparing a world where everyone receives an intervention
versus a world where everyone receives a control---the so-called
average treatment effect. For multi-valued exposures, most methods
study the outcome expectation under a series of hypothetical worlds
corresponding to assigning the same value of the exposure to all units
in the population---the so-called dose-response
curve. 
While informative, a dose-response analysis presents some challenges.
First, for some exposures (e.g., physical activity) it is hard to
conceive an intervention that would set them statically (e.g., make
everyone exercise 30 minutes), even in principle. This presents a
problem for dose-response effects because one cannot reasonably
implement an intervention that would bring about such effects in the
real world. Second, summarizing the infinite-dimensional dose-response
curve often requires restrictive and arbitrary parametric
assumptions. These assumptions, frequently encoded in marginal
structural models, are typically wrong, and the models can be hard to
interpret under model misspecification
\citep{Neugebauer2007419}. Third, non-parametric approaches to
estimation of the causal dose-response curve cannot achieve
$n^{1/2}$-consistency, because the parameter is not pathwise
differentiable. For example, the estimators of \cite{kennedy2017non}
converge at the $(nh)^{1/2}$-rate (where $h$ is a bandwidth that
decreases with sample size, giving the familiar $n^{-2/5}$ rate under
twice-differentiability)
. Fourth, the
effects are not identified when some units have a zero chance to
receive some of the exposure levels under consideration, a situation
known as violations to the \textit{positivity assumption}. Positivity
violations are likely present with most continuous and multi-valued
exposures, and are exacerbated when the exposures are measured at
multiple time points.

As a solution to the above limitations, the causal inference
literature has considered alternative definitions of causal effects,
allowing for hypothetical worlds where the exposures can depend on
characteristics of the unit (dynamic regimes), where the
post-intervention distribution is a random draw from a given but
possibly unknown distribution (stochastic interventions), and where
the post-intervention exposure may be a random or deterministic
function of the observed treatment value \citep[e.g.,
][]{stock1989nonparametric,robins2004effects}. In this paper we adopt
the latter approach. In this context, we have previously studied
interventions that shift the exposure distribution \citep{Diaz12} for
single time point studies. \cite{Haneuse2013} generalized our previous
interventions and introduced the name \textit{modified treatment
  policy (MTP)}, which we adopt in this manuscript. MTPs can yield
estimates with a familiar interpretation as the expected change in
mean response for a given change in the exposure of interest---without
the need to impose unrealistic parametric assumptions such as
linearity in the causal model. In our motivating example, we estimate
the effect on 14-day survival of an increase of 50 units in the
arterial partial pressure of oxygen to fraction of inspired oxygen
(P/F) ratio, where the only patients with respiratory failure (P/F
ratio $< 300$) receive the
intervention. 

For the longitudinal case, \cite{robins2004effects} introduced
\textit{dynamic regimes depending on the natural value of treatment},
where the \textit{natural value of treatment} is the value that
treatment would take at time $t$ if the intervention was discontinued
right before time $t$. \cite{richardson2013single} and
\cite{young2014identification} formalized the effect definition and
showed conditions under which the effects are identified by the
extended g-formula of \cite{robins2004effects}. A key finding of these
papers is that the sequential randomization assumption required for
identifying effects that depend on the natural value of treatment is
stronger than that required for identification of the effect of
dynamic regimes. Although \cite{richardson2013single} and
\cite{young2014identification} use the same name \textit{dynamic
  regimes depending on the natural value of treatment} to refer to the
interventions, the definitions given in the papers are different. In
this article we use the name \textit{longitudinal modified treatment
  policies (LMTP)} to refer to the definition of
\cite{richardson2013single}, as the interventions in this work are a
natural extension of the definition of \cite{Haneuse2013} for single
time point studies. 

We present a novel alternative expression of the extended g-formula in
terms of sequential regressions.  We show conditions under which the
sequential regression formula allows the generalization of estimators
for dynamic regimes to the case of LMTPs. We also introduce a
stochastic intervention where instead of setting the exposure equal to
the LMTP, we set it to a random draw from the LMTP distribution. We
call this intervention LMTP stochastic intervention
(LMTP-SI). 
\cite{young2014identification} show that the sequential randomization
assumption required for identification of LMTP-SI is identical to that
required for identification of the effect of dynamic
interventions. 
Thus, the LMTP functional can
still be interpreted as a causal effect even under the standard
identifiability assumptions required for dynamic regimes.

We propose four different estimators for the LMTP sequential
regression functional. The first estimator is a simple extension of
the inverse parametric probability weighted (IPW) of
\cite{young2014identification}; the second is a simple extension of
the g-computation estimator presented by \cite{robins2004effects} and
\cite{taubman2009intervening}. IPW and g-computation estimation
require estimating nuisance parameters related to the conditional
density of treatment and outcome regressions conditional on the
history of a unit. When these nuisance parameters are estimated within
pre-specified parametric models, the LMTP estimators are
asymptotically Gaussian and the Delta method or the bootstrap are
guaranteed to yield valid asymptotic p-values and confidence
intervals. However, whenever the nuisance estimators are
data-adaptive, such as when model selection is performed, the
asymptotic distribution of the IPW and g-computation estimators is
generally unknown, making it difficult to perform hypothesis tests and
quantify the uncertainty around the estimate. We thus develop two
additional estimators that overcome this issue.



Our third and fourth estimation strategies have roots in
semi-parametric estimation theory \citep[e.g.,][]{mises1947asymptotic,
  vanderVaart98, robins2009quadratic, Bickel97}, in the theory for
doubly robust estimation using estimating equations
\citep{Robins00,Robins&Rotnitzky&Zhao94,vanderLaan2003,Bang05}, and in
the targeted learning framework \citep{vdl2006targeted,
  vanderLaanRose11, vanderLaanRose18}. Central to this theory is the
study of the efficient influence function (EIF) or canonical gradient,
which characterizes the efficiency bound of the LMTP functional and
allows the development of estimators under slow convergence rates for
the nuisance parameters involved \citep{robins2009quadratic}. We
derive the EIF for LMTPs under the assumption that the LMTP does not
depend on the data generating mechanism. We show that this EIF has a
similar structure to the EIF for dynamic interventions under a
differentiability and invertibility condition on the LMTP. We then
show that, under these conditions, the EIF allows for multiply robust
estimation. This is a surprising fact that should not be expected for
general interventions that depend on the natural value of treatment,
since the parameter functional depends on both the outcome regression
and the exposure mechanism. We exploit the similarity of the EIF to
generalize some estimators for dynamic intervention available in the
literature. In particular, we develop a targeted minimum loss-based
estimator (TMLE), which is a natural extension of the estimator of
\cite{van2012targeted}. We show that the TMLE is multiply robust
consistent in the sense that it allows for consistent estimation under
$\tau+1$ configurations of consistent estimation of the nuisance
parameters, where $\tau$ is the number of time points under
consideration. We then develop a sequentially doubly robust (SDR)
estimator based on multiply robust unbiased transformations by
generalizing the estimators of \cite{luedtke2017sequential} and
\cite{rotnitzky2017multiply}, which are themselves related to ideas
dating back to \cite{rubin2007doubly} and
\cite{tchetgen2009commentary}. \cite{luedtke2017sequential} refer to
this estimator as \textit{SDR via doubly robust transformations}; we
adopt this name and shorten it to SDR for simplicity.
The SDR estimator is expected to be consistent under $2^\tau$
configurations of consistent estimation of the nuisance
parameters. This is in complete analogy to the $(\tau+1)$- and
$2^\tau$-multiply robust consistency results discussed by
\cite{luedtke2017sequential}, \cite{rotnitzky2017multiply}, and
\cite{molina2017multiple} for the case of a dynamic intervention.  We
use cross-fitting to obtain $n^{1/2}$-convergence of our estimators
while avoiding entropy conditions that may be violated by data
adaptive estimators of the nuisance parameters \citep{zheng2011cross,
  chernozhukov2018double}.  Finally, we note that since the LMTP
parameter studied in this paper generalizes dynamic interventions for
longitudinal studies, our methods and software
can also be used for estimation of those parameters.
\section{Notation and definition of causal effects}\label{sec:nota}
Let $Z_1,\ldots, Z_n$ denote a sample of i.i.d. observations with
$Z=(L_1, A_1, L_2, A_2, \ldots, L_\tau, A_\tau, Y)\sim \P$, where
$L_t$ denotes time-varying covariates, $A_t$ denotes a vector of
intervention variables such as treatment and/or censoring status, and
$Y=L_{\tau+1}$ denotes an outcome such as survival at the end of study
follow-up. We let $\P f = \int f(z)\dd \P(z)$ for a given function
$f(z)$. We use $\Pn$ to denote the empirical distribution of
$Z_1,\ldots\,Z_n$, and assume $\P$ is an element of the nonparametric
statistical model defined as all continuous densities on $Z$ with
respect to a dominating measure $\nu$. We let $\E$ denote the
expectation with respect to $\P$, i.e.,
$\E\{f(Z)\} = \int f(z)\dd\P(z)$. We also let $||f||^2$ denote the
$L_2(\P)$ norm $\int f^2(z)\dd\P(z)$. We use
$\bar X_t = (X_1,\ldots, X_t)$ to denote the history of a variable,
use $\underline X_t = (X_t,\ldots, X_\tau)$ to denote the future of a
variable, and use $H_t = (\bar A_{t-1}, \bar L_t)$ to denote the
history of all variables up until just before $A_t$. For the complete
history of a random variable, we simplify $\bar X_\tau$ as $\bar
X$. We let $\g_t(a_t \mid h_t)$ denote the probability density
function of $A_t$ conditional on
$H_t=h_t$
. We use
calligraphic font to denote the support of a random variable, e.g.,
$\mathcal A_t$ denotes the support of $A_t$.



We formalize the definition of the causal effects using a
non-parametric structural equation model
\citep{Pearl00}. Specifically, for each time point $t$, we assume the
existence of deterministic functions $f_{L_t}$, $f_{A_t}$, and $f_Y$
such that $L_t=f_{L_t}(A_{t-1}, H_{t-1}, U_{L,t})$,
$A_t=f_{A_t}(H_t, U_{A,t})$, and $Y=f_Y(A_\tau, H_\tau, U_Y)$. Here
$U=(U_{L,t}, U_{A,t}, U_Y:t\in \{1,\ldots,\tau\})$ is a vector of
exogenous variables, with unrestricted joint distribution. Sufficient
assumptions to identify the effects we discuss will be given in \S
\ref{sec:iden}. LMTP effects can be defined in terms of hypothetical
interventions where the equation $A_t=f_{A_t}(H_t, U_{A,t})$ is
removed from the structural model, and the exposure is assigned as a
new random variable $A_t^\d$. An intervention that sets the exposures
up to time $t-1$ to $\bar A_{t-1}^\d$ generates counterfactual
variables
$L_t(\bar A_{t-1}^\d) = f_{L_t}(A_{t-1}^\d, H_{t-1}(\bar A_{t-2}^\d),
U_{L,t})$ and
$A_t(\bar A_{t-1}^\d) = f_{A_t}(H_t(\bar A_{t-1}^\d), U_{A,t})$, where
the counterfactual history is defined recursively as
$H_t(\bar A_{t-1}^\d) = (\bar A_{t-1}^\d, \bar L_t(\bar
A_{t-1}^\d))$. The variable $A_t(\bar A_{t-1}^\d)$ is referred to as
the \textit{natural value of treatment} \citep{richardson2013single,
  young2014identification}, and represents the value of treatment that
would have been observed at time $t$ under an intervention carried out
up until time $t-1$ but discontinued thereafter. An intervention where
all the treatment variables up to $t=\tau$ are intervened on generates
a counterfactual outcome
$Y(\bar A^\d)=f_Y(A_\tau^\d, H_\tau(\bar A_{\tau-1}^\d), U_Y)$. Causal
effects will be defined in terms of the distribution of this
counterfactual. We discuss two types of effects: longitudinal modified
treatment policies (LMTPs), and LMTP stochastic interventions
(LMTP-SI). Both effects are defined in terms of a user-given
function $\d(a_t, h_t)$ that maps a treatment value $a_t$ and a
history $h_t$ into a new exposure value. For fixed values $\bar a_t$,
$\bar l_t$, we recursively define $a_t^\d=\d(a_t, h^\d_t)$, where
$h^\d_t=(\bar a_{t-1}^\d, \bar l_t)$. The LMTPs that we study are thus
defined as follows:

\begin{definition}[Longitudinal modified treatment policies (LMTP)]\label{def:mtpn}
  An intervention $A_t^\d$ is said to be a longitudinal modified
  treatment policy if it is defined as
  $A^\d_t = \d(A_t(\bar A_{t-1}^\d), H_t(\bar A_{t-1}^\d))$ for a
  user-given function $\d$.  LMTP causal effects are defined as
  contrasts between the distribution of counterfactual outcomes
  $Y(\bar A^\d)$ under different functions $\d$.
\end{definition}

This definition is a longitudinal generalization of the modified
treatment policies defined by \cite{Haneuse2013} for a single time
point, because at $t=1$ the factual and natural values of treatment
coincide. For $\tau>1$, the modified treatment policies in
Definition~\ref{def:mtpn} are a particular case of the interventions
that depend on the natural value of treatment first discussed by
\cite{robins2004effects} and formalized by \cite{richardson2013single}
and \cite{young2014identification}.

To illustrate some of our methods and ideas, we will consider the
following important examples of regimes $\d$.

\begin{example}[Threshold LMTP]\label{ex:threshold} We revisit a problem posed by
  \cite{taubman2009intervening}, in which we are interested in
  assessing the effect of exercising at least 30 minutes a day on the
  risk of coronary heart disease (CHD). Let $A_t$ denote the minutes
  exercised by each study participant, let $Y$ denote an indicator of
  CHD by the end of the study, and let $L_t$ denote confounders such
  as comorbidities and lifestyle variables. Let
  $\d(a_t, h_t) = \one(a_t \geq 30)a_t + \one(a_t < 30)30$. At time
  point $t=1$, we define the intervention as $A^\d_1 = \d(A_1, H_1)$,
  which would set the number of minutes exercised to 30 if the
  participant exercised less than 30, and would leave it unchanged
  otherwise.  At time points $t>1$, we are interested in setting the
  amount of physical activity to 30 if the participant's natural value
  of treatment at time $t$ is smaller than 30, and set it to the
  natural value of treatment otherwise. That is, we define
  $A_t^\d=\d(A_t(\bar A_{t-1}^\d), H_t(\bar A_{t-1}^\d))$.
\end{example}

\begin{example}[Shift LMTP]
  Let $A_t$ denote a continuous exposure, such as a drug dose or a
  physiological measurement such as P/F ratio that can be modified
  through intervention.  To define this intervention, assume that
  $A_t$ is supported as $\P(A_t \leq u_t(h_t)\mid H_t = h_t)=1$ for
  some $u_t$. Then, for a user-given value $\delta$, we let
  \begin{equation}\label{eq:defdshift}
    \d(a_t,h_t)=
    \begin{cases}
      a_t + \delta & \text{if } a_t \leq u_t(h_t) - \delta \\
      a_t & \text{if } a_t > u_t(h_t) - \delta.
    \end{cases}
  \end{equation} Then we define
  $A_t^\d=\d(A_t(\bar A_{t-1}^\d), H_t(\bar A_{t-1}^\d))$. This intervention was first
  introduced in the context of a single time point by~\cite{Diaz12},
  and has been further discussed by~\cite{diaz2018stochastic} and
  \cite{Haneuse2013}, and by \cite{diaz2020} in the context of
  mediation. This intervention considers hypothetical worlds
  in which the natural exposure at time $t$ is increased by a
  user-given value $\delta$, whenever such increase is feasible for
  a unit with history $H_t(A_{t-1}^\d)$. In our motivating example,
  we assess the effect on mortality of an intervention that would
  increase a patient's P/F ratio by 50 units for patients with
  acute respiratory failure (P/F ratio $<$ 300). Alternatively, we
  could define a multiplicative shift as $\d(a_t,h_t)= a_t\delta(h_t)$
  for a shift function $\delta(h_t)$ that may depend on the history $h_t$.
\end{example}

  

In some cases, an alternative intervention may be of
interest. Consider for example the implementation of a public health
intervention to encourage people to exercise more. In this case, one
could be interested in a post-intervention exposure where the
distribution of time exercise has shifted, but that does not
necessarily set each individual's exposure to
$A_t^\d=\d(A_t(\bar A_{t-1}^\d), H_t(\bar A_{t-1}^\d))$ with $\d$ defined in
(\ref{eq:defdshift}). This intervention is different from a modified
treatment policy in that we are not necessarily interested in shifting
an individual's exposure. Instead, we  wonder what would have happened in a
hypothetical world where each unit's exposure was a random draw from a
shifted distribution, representing a behavior shift at the population
level. A LMTP stochastic intervention is then formalized as:

\begin{definition}[LMTP stochastic intervention (LMTP-SI)]
  Let $\G^\d_t$ denote the distribution of $\d(A_t, H_t)$ conditional
  on $H_t$. A longitudinal modified treatment policy stochastic
  intervention $\bar Q^{\d}$ is defined as a sequence of random draws
  from $\G^\d_t: t=1,\ldots,\tau$. LMTP-SI Causal effects are defined
  as contrasts between the distribution of counterfactual outcomes
  $Y(\bar Q^\d)$ under different functions $\d$.
\end{definition}

In the examples discussed so far we consider interventions where the
function $\d(a_t,h_t)$ only depends on values of the exposure $a_t$
and the history $h_t$. We note that the definition of
\cite{richardson2013single} allows for random regimes by letting the
functions $\d(a_t,h_t,\varepsilon_t)$ depend on a randomizer
$\varepsilon_t$. It is easy to imagine such settings in practice. For
example, \cite{robins2004effects} propose to study the effect on
coronary heart disease of interventions where a random half of smokers
quit smoking forever. The methods that we present next also allow for
this type of random intervention. However, we require some assumptions
on the randomizer $\varepsilon_t$, namely that (i) it is drawn
independently across units and independently of $U$, and (ii) its
distribution does not depend on $\P$. Under these two assumptions, and
with the aim of simplifying notation, we assume without loss of
generality that the time varying vector $L_t$ contains the randomizer
$\varepsilon_t$. Another example of such a randomized intervention is
given below.

\begin{example}[Incremental propensity score
  interventions]\label{ex:ipsi}
  \cite{kennedy2018nonparametric} proposed an intervention for binary
  exposures where $A^\d_t$ is a draw from a Bernoulli distribution
  with shifted propensity score
  given by
  \[\g_t^\d(1\mid h_t) = \frac{\delta \g_t(1\mid h_t)}{\delta
      \g_t(1\mid h_t) + 1 - \g_t(1\mid h_t)}.\]
  Let $\varepsilon_t$ denote a random draw from a uniform distribution
  in the interval $(0,1)$. Define the intervention as $\d_t(a_t, h_t) =
  \one\{\varepsilon_t < \g_t^\d(1\mid h_t)\}$.
\end{example}

In what follows we are concerned with identification and
non-parametric estimation of the causal parameters
\[\thetalmtp = \E\{Y(\bar A^\d)\},\text{ and }\thetalsi = \E\{Y(\bar Q^\d)\},\]
where $Y(\bar A^\d)$ and $Y(\bar Q^\d)$ are defined above. The
difference between LMTP and LMTP-SI is subtle but important, and leads
to different assumptions required for identification, as we will see
in the next section.

\section{Identification of causal effects}\label{sec:iden}

The first step in developing estimators for $\thetalmtp$ and
$\thetalsi$ is to derive an identification result that allows us to
write these causal parameters as a function of only the distribution
$\P$ of the observed data $Z$. Sufficient assumptions to identify
$\thetalmtp$ and $\thetalsi$ were first given by
\cite{richardson2013single} and \cite{young2014identification}. We
first present sufficient assumptions for identification under the
assumed NPSEM, and then discuss their implications in several
scenarios.
\begin{assumption}[Positivity]\label{ass:support}
  If $(a_t,h_t)\in \supp\{A_t,H_t\}$ then
  $(\d(a_t,h_t),h_t)\in \supp\{A_t,H_t\}$ for $t\in\{1,\ldots,\tau\}$.
\end{assumption}
\begin{assumption}[Standard sequential randomization]\label{ass:exch1}
  $U_{A,t}\indep \underline U_{L, {t+1}}\mid H_t$ for all  $t\in\{1,\ldots,\tau\}$.
\end{assumption}

\begin{assumption}[Strong sequential randomization]\label{ass:exch}
  $U_{A,t}\indep (\underline U_{L, {t+1}}, \underline U_{A, t+1}) \mid H_t$ for all  $t\in\{1,\ldots,\tau\}$.
\end{assumption}
Assumption \ref{ass:support} is equivalent to the assumption presented
in \cite{young2014identification}, and simply states that the
distribution of interest is supported in the data. Consider our
motivating example of the effect of P/F ratio on survival on ICU
patients. Under no loss-to-follow-up, this assumption states that if
it is possible to find a patient with history $h_t$ who has a P/F
ratio of $a_t$ at time $t$, then it is also possible to find a patient
with history $h_t$ who has a P/F ratio of $\d(a_t,h_t)$. We note that
this assumption may be enforced by definition of $\d$ if sufficient
information is available about the conditional support of $a_t$
conditional on $h_t$. Furthermore, if $a_t$ is multivariate and
includes missingness or censoring indicators, then Assumption
\ref{ass:support} also states that for every observed history $h_t$
there is a probability greater than zero of observing a patient who is
not lost-to-follow-up at time $t$.  Assumption \ref{ass:exch1} is
standard for the identification of dynamic regimes, and is satisfied
if all the common causes of the intervention variable $A_t$ and
$L_s:s>t$ are measured and recorded in $H_t$. Assumption
\ref{ass:exch} is stronger than Assumption \ref{ass:exch1}, and is
satisfied if all common causes of the intervention variable $A_t$ and
$(A_s, L_s):s>t$ are measured and recorded in $H_t$. Assumption
\ref{ass:exch} is similar in nature to the independence assumption
required by \cite{richardson2013single} (see Theorem 31 in that
reference).

We have the following identification theorem, which allows us to
compute the parameters $\thetalmtp$ and $\thetalsi$ as a function only
of the observed data distribution.

\begin{theorem}[Identification of the effect of LMTPs]\label{theo:iden}
  Set $\Q_{\tau+1}= Y$. For $t=\tau,\ldots,1$, recursively define
  \begin{equation}
    \Q_t:(a_t, h_t) \mapsto \E\left[\Q_{t+1}(A_{t+1}^\d, H_{t+1})\mid
      A_t=a_t, H_t=h_t\right],\label{eq:defQ}
  \end{equation}
  and define $\theta = \E[\Q_1(A_1^\d, L_1)]$. Then we have:
  \begin{enumerate}[label=(\roman*)]
  \item Under Assumptions \ref{ass:support} and \ref{ass:exch1}, $\thetalsi$
    is identified as $\theta$. 
  \item Under Assumptions \label{theo:swig} \ref{ass:support} and
    \ref{ass:exch}, $\thetalmtp$ is identified as $\theta$.
  \end{enumerate}
\end{theorem}
The identification expression given in Theorem~\ref{theo:iden} in
terms of sequential regressions is an alternative expression of the
extended g-formula of \cite{robins2004effects} and
\cite{richardson2013single}. 
In addition to allowing for a variety of interesting interventions,
this setup allows for a variety of data structures. In particular, it
can handle loss-to-follow-up, survival analysis, and missing exposures
as follows. Let $A_t=(A_{1,t},A_{2,t})$, where $A_{1,t}$ denotes the
exposure at time $t$, and $A_{2,t}$ is equal to one if the unit
remains uncensored at time $t+1$ and zero otherwise. Assume monotone
loss-to-follow-up so that $A_{2,t}=0$ implies $A_{2,k}=0$ for all
$k>t$, in which case $(L_k,A_{1,k})$ and $Y$ become degenerate for
$k > t$. In this case we could be interested in a hypothetical world
in which there is no loss-to-follow-up and the exposure $A_{1,t}$ is
shifted as in (\ref{eq:defdshift}) or any other intervention of
interest. In particular, we can define
$A^\d_t = (\d\{A_{t,1}(\bar A_{t-1, 1}^\d),H_t(\bar A_{t-1, 1}^\d)\},
1)$. Time-varying outcomes can be incorporated by letting
$L_t=(X_t, Y_{t-1})$, where $X_t$ denotes the time-varying covariates
of interest and $Y_t$ denotes the time-varying outcome, and letting
$Y=Y_\tau$. If a prior time point $k$ is of interest, then we let
$Y=Y_k$ and truncate the sequence at $\tau=k$. Time-to-event analysis
may be performed by letting $Y_t$ denote an indicator that a unit is
event free at time $t$, and letting $A_{2,t}$ denote an indicator that
the unit is uncensored at time $t+1$. This definition of the random
variables imposes restrictions on the functions $\Q_t$ and $\r_t$,
which must be taken into account in estimation. For example, we know
by definition that $\Q_t(a_t, h_t)=1$ for $h_t$ such that $y_s=1$ for
any $s<t$.
\section{Optimality theory}

Thus far we have derived a novel sequential regression formula that
identifies the effect of an LMTP. We now turn our attention to a
discussion of efficiency theory for its estimation in the
nonparametric model. The \textit{efficient influence function} (EIF)
is a key object in semi-parametric estimation theory, as it
characterizes the asymptotic behavior of all regular and efficient
estimators \citep{Bickel97}. Knowledge of the EIF has important
practical implications. First, the EIF is often useful in constructing
locally efficient estimators. Second, the EIF estimating equation
often enjoys desirable properties such as multiple robustness, which
allows for some components of the data distribution to be
inconsistently estimated while preserving consistency of the
estimator. Third, asymptotic analysis of estimators constructed using
the EIF often yields second-order bias terms, which require slow
convergence rates (e.g., $n^{-1/4}$) for the nuisance parameters
involved, thereby enabling the use of flexible regression techniques
in estimating these quantities. Before we proceed developing such
theory, note that it is not possible to construct $n^{1/2}$-consistent
estimators of $\theta$ for the threshold intervention in
Example~\ref{ex:threshold} where
$\d(a_t,l_t)= a_t \one(a_t\leq \delta) + \delta I(a_t > \delta)$. This
is because the parameter is not pathwise differentiable. Intuitively,
inspection of the parameter definition for $\tau=1$ yields the reason
for the lack of $n^{1/2}$-estimability:
\begin{align*}
  \theta&=\E\{\Q(A \one(A\leq \delta) + \delta \one(A >
          \delta), L)\}\\
        &=\E\{\Q(A,L) \one(A\leq \delta)\} + \E\{\Q(\delta, L)\one(A >
          \delta)\}.
\end{align*}
The term $\E\{\Q(\delta, L)\one(A > \delta)\}$ in this expression
involves estimation of the causal effect of a static intervention
setting a continuous exposure to $A=\delta$. Efficient estimation
theory is not available for estimation of such parameters in the
non-parametric model \citep{Bickel97}, since all possible gradients of
the pathwise derivative would necessarily involve a Dirac delta
function at $\delta$. An alternative approach to overcome this issue
is to redefine the regime $\d$ so that the parameter becomes pathwise
differentiable. Such approach is taken by \cite{diaz2013assessing};
the interested reader is encouraged to consult the original research
article. In this article we avoid this problem by only considering
interventions that satisfy the following assumption, which is a
straightforward generalization of the assumption of \cite{Haneuse2013}
for $\tau=1$.
\begin{assumption}[Piecewise smooth invertibility for continuous exposures]\label{ass:inv}
  For each $h_t$, assume that the support of $A_t$ conditional on
  $H_t=h_t$ may be partitioned into subintervals
  ${\cal I}_{t,j}(h_t):j = 1, \ldots, J_t(h_t)$ such that
  $\d(a_t, h_t)$ is equal to some $\d_j(a_t, h_t)$ in
  ${\cal I}_{t,j}(h_t)$ and $\d_j(\cdot,h_t)$ has inverse function
  $\b_j(\cdot, h_t)$ with derivative $\b_j'(\cdot, h_t)$ with
  respect to $a_t$.
\end{assumption}
Define
\begin{equation}\label{eq:gdelta}
  \g_t^\d(a_t \mid h_t) =
  \sum_{j=1}^{J_t(h_t)} \one_{t, j} \{\b_j(a_t, h_t), h_t\} \g_t\{\b_j(a_t, h_t)\mid h_t\}
  |\b_j'(a_t,h_t)|,
\end{equation}
where $\one_{t,j} \{u, h_t\} = 1$ if $u \in {\cal I}_{t, j}(h_t)$ and
$\one_{t,j} \{u, h_t\} = 0$ otherwise. Under Assumption \ref{ass:inv},
it is easy to show that the p.d.f. of $A_t^\d$ conditional on the
history $h_t$ is $\g_t^\d(a_t \mid h_t)$. In the case of equation
(\ref{eq:defdshift}) the post-intervention p.d.f. becomes
\begin{equation*}
  \g_t^\d(a_t\mid h_t) = \g_t(a_t-\delta\mid h_t) \one\{a_t < u_t(h_t)\} +
  \g_t(a_t\mid h_t) \one\{a_t +\delta \geq u_t(h_t)\},
\end{equation*}
which shows that piecewise smoothness is sufficient to handle
interventions such as (\ref{eq:defdshift}) which are not smooth in the
whole range of the exposure. 

For discrete exposure variables, we let
\[\g^\d_t(a_t\mid h_t) = \sum_{s_t\in \mathcal A_t}\one\{\d(s_t, h_t)=a_t\}\g_t(s_t\mid
h_t).\] Assumption \ref{ass:inv} and expression (\ref{eq:gdelta})
ensure that we can use the change of variable formula when computing
integrals over $\mathcal A_t$ for continuous exposures. This is useful
for studying properties of the parameter and estimators we propose.

Efficiency theory in this paper focuses on functions $\d$ that do not
depend on $\P$ (recall that the function is deterministic but allowed
to take a randomizer as argument). Assumption \ref{ass:inv} together
with this assumption will ensure that the efficient influence function
of $\theta$ for LMTPs has a structure similar to the influence
function for the effect of dynamic regimes. This yields two important
advantages for estimation. First, the structure of the EIF allows for
multiply robust estimation, which is not generally possible for random
regimes  $\d$
that depend on $\P$. See \cite{diaz2013assessing} and
\cite{kennedy2018nonparametric} (Example~\ref{ex:ipsi} above) for
examples. Second, this similarity will allow us to generalize existing
estimators for dynamic regimes. In what follows it will be useful to
define the density ratio
\[\r_t(a_t, h_t) = \frac{\g_t^\d(a_t\mid
    h_t)}{\g_t(a_t\mid h_t)},\]
and the function 
\begin{equation}
  \label{eq:eifgamma}
  \D_t : z\mapsto
  \sum_{s=t}^\tau\left(\prod_{k=t}^s \r_k(a_k, h_k)\right)\{\Q_{s+1}(a_{s+1}^\d, h_{s+1}) -
  \Q_s(a_s, h_s)\} + \Q_t(a_t^\d, h_t)
\end{equation}
for $t=\tau,\ldots,1$. When necessary, we use the notation
$\D_t(z;\eta)$ or $\D_t(z;\underline\eta_t)$ to highlight the
dependence of $\D_t$ on
$\underline\eta_t=(\r_t,\Q_t,\ldots,\r_\tau, \Q_\tau)$
. We also use $\eta$ to denote
$(\r_1,\Q_1,\ldots,\r_\tau, \Q_\tau)$, and define
$\D_{\tau+1}(Z;\eta) = Y$.

\begin{theorem}[Efficient Influence Function]\label{theo:eif}
  Assume one of:
  \begin{enumerate*}[(i)]
  \item $A_t$ is a discrete random variable for all $t$, or
  \item $A_t$ is a continuous random variable and the modified
    treatment policy $\d$ satisfies Assumption~\ref{ass:inv}.
  \end{enumerate*}
 \textcolor{black}{Assume that $\d$ does not depend on $\P$.}
  The efficient influence function for estimating
  $\theta = \E[\Q_1(A^\d,L_1)]$ in the non-parametric model is given
  by $\D_1(Z)-\theta$.
\end{theorem}
Note that if $\tau=1$, then $\D_1(Z)-\theta$ reduces to the efficient
influence function for a single time point intervention presented by
\cite{Diaz12}, equal to
$\r(A, L)[Y - \Q(A,L)] + \Q(\d(A,L), L) -\theta$. For the case
$\tau=1$, the efficiency bound equals
$\var\{\D_1(Z)-\theta\}=\E[\r^2(A,L)\var(Y\mid A,L)] + \E[\Q(\d(A, L),
L)-\theta]^2$, which depends on three features of the data
distribution and intervention: (i) the conditional variance of the
outcome $\var(Y\mid A,L)$, (ii) the amount of treatment effect
heterogeneity $\E[\Q(\d(A, L), L)-\theta]^2$, and (iii) the extent to
which the intervention $\d$ differs from the observed regime, as
measured by the density ratio $\r(A,L)$. Furthermore, interventions
that exert large exposure density changes in areas of high outcome
variability are expected to yield larger efficiency bounds, since
$\r^2(A,L)$ and the variance $\var(Y\mid A,L)$ are positively
correlated in such cases.

In the following, we let
$\eta'=(\r_1',\Q_1',\ldots,\r_\tau', \Q_\tau')$ denote some value of
$\eta$. This value will typically represent the probability limit of a
given estimator $\hat\eta$. The efficient influence function satisfies
the following property, which will be crucial to establish consistency
of some estimators of $\theta$ under multiple robustness assumptions,
and to construct estimators of $\theta$ under slow convergence rates
for estimation of $\r_t$ and $\Q_t$.

\begin{lemma}[First-order parameter approximation]\label{theo:sdr}
  Let
  \[C_{t, s}' = \prod_{r=t+1}^{s-1}\r_r'(A_r,H_r).\]
  For each $t\in\{0,\ldots,\tau-1\}$, and for any $\eta'$, define the second order error
  term {\footnotesize
    \begin{multline}
      \rem_t(a_t, h_t;\eta') =\\
      \sum_{s=t+1}^\tau\E\left[C_{t,s}'\{\r_s'(A_s,
        H_s) - \r_s(A_s, H_s)\}\{\Q_s'(A_s, H_s) -
        \Q_s(A_s, H_s)\}\,\,\bigg|\,\, A_t=a_t,H_t=h_t \right],\label{eq:rem}
    \end{multline}
  } where for $t=0$ the conditioning event is the null set, and for
  $t=\tau$ we let $\rem_\tau(a_\tau, h_\tau;\eta')=0$. Under
  Assumption~\ref{ass:inv} we have
  \begin{equation}
    \Q_t(a_t,h_t) =
    \E\big[\D_{t+1}(Z;\eta')\mid A_t=a_t,H_t=h_t\big] + \rem_t(a_t, h_t;\eta').\label{eq:fo}
  \end{equation}
\end{lemma}

This lemma is analogous to Lemma 1 in \cite{luedtke2017sequential} and
Lemma 2 in \cite{rotnitzky2017multiply} for the standard g-formula for
dynamic regimes. Furthermore, this lemma shares important connections
to the von Mises-type expansions used in some of the semi-parametric
inference literature \citep[e.g.,][]{mises1947asymptotic,
  vanderVaart98, robins2009quadratic}. For $t=0$, inspection of this
lemma teaches us that it is possible to construct a consistent
estimator of $\theta$ by averaging $\D_1(Z_i;\hat\eta)$ across the
sample, where $\hat\eta$ is an estimator such that
$\rem_0(\hat\eta)=o_\P(1)$. The latter consistency can be achieved
under the condition that for each $t$, either $\r_t$ or $\Q_t$ can be
estimated consistently. This robustness is an interesting property in
light of the fact that the parameter $\theta$ depends on both $\r_t$
and $\Q_t$ for all
$t$. 

\section{Estimation and statistical inference}\label{sec:estima}

In this section we assume that preliminary estimators $\hat\r_t$, and
$\hat\Q_t$ are available. These estimators may be obtained from
flexible regression techniques such as support vector machines,
regression trees, boosting, neural networks, splines, or ensembles
thereof \citep{Breiman1996, vanderLaan&Polley&Hubbard07}. As
previously discussed, the consistency of these estimators will
determine the consistency of our estimators of the parameter
$\theta$. In particular, $\Q_t:t=1,\ldots,\tau$ may be estimated as
follows. Start by running a preferred regression algorithm of
$\Q_{\tau+1,i} = Y_i$ on $(A_{\tau,i},H_{\tau,i})$. Then evaluate the
estimator $\hat\Q_\tau$ at $(A_{\tau,i}^\d,H_{\tau,i})$, i.e., compute
the prediction $\hat\Q_\tau(A_{\tau,i}^\d,H_{\tau,i})$. Use this
prediction as the pseudo-outcome in a regression on
$(A_{\tau-1,i}, H_{\tau-1,i})$, to obtain an estimate
$\hat\Q_{\tau-1}$. Compute the pseudo-outcome
$\hat\Q_{\tau-1}(A_{\tau-1,i}^\d,H_{\tau-1,i})$ and iterate the
process until obtaining an estimate $\hat\Q_1$. For this and other
estimators presented below it will be necessary to specify a
regression algorithm to estimate conditional expectations. In this
paper we advocate for the use of data-adaptive regression methods,
which avoid reliance on tenuous parametric assumptions that can
invalidate the conclusions of an otherwise well-designed and conducted
study. In particular, we propose to use an ensemble regression
algorithm known as the Super Learner
\citep{vanderLaan&Polley&Hubbard07}, which builds a convex combination
of regression algorithms in a user-given library, with weights chosen
to minimize the cross-validated prediction error. 
While estimation methods for
conditional expectations abound in the statistics and machine learning
literature, data-adaptive methods to estimate a multivariate density
ratio such as $\r_t$ are scarce. In \S~\ref{sec:densratio} below we
present a method in which we recast the density ratio estimation
problem as a classification problem based on $2n$ observations. Once
the problem is recast in this way, any classification method from the
statistical learning literature (such as Super Learning) may be used
to estimate the density ratio $\r_t$.

We start this section by discussing two simple estimators: inverse
probability weighting and g-computation or substitution
estimators. These estimators cannot generally achieve
$n^{1/2}$-consistency under data-adaptive estimation of the nuisance
parameters in a non-parametric model. We then present a targeted
minimum loss-based estimator which is locally efficient,
$n^{1/2}$-consistent, and $\tau+1$-multiply robust consistent, under
assumptions. We then present a sequential regression estimator which
has the same properties but has the additional advantage that it is
$2^\tau$-multiply robust consistent.

\subsection{Substitution and inverse probability weighted estimators}

The substitution estimator simply uses the preliminary estimator of
$\Q_t$ described above along with the recursive definition in equation
(\ref{eq:defQ}):
\[\thetasub = \frac{1}{n}\sum_{i=1}^n\hat\Q_1(A_{1,i}^\d,L_{1,i}).\] The IPW
estimator is based on the observation that, under Assumption
\ref{ass:inv}, we have the following alternative representation of the
parameter of interest (see Lemma~\ref{lemma:altrep} in the
supplementary materials):
\[\theta = \E\left[\left(\prod_{t=1}^\tau \r_t(A_t, H_t)\right) Y\right].\]
Thus, for estimators $\hat\r_t$ of $\r_t$, we define
\[\thetaipw = \frac{1}{n}\sum_{i=1}^n \left(\prod_{t=1}^\tau \hat
    \r_t(A_{t,i}, H_{t,i})\right) Y_i.\] This estimator is an
extension of the estimator proposed by \cite{young2014identification},
where we allow the use of machine learning methods to estimate the
density ratio (see \S\ref{sec:densratio}).

If $\Q_t$ and $\r_t$ are estimated within pre-specified parametric
models, then, by the delta method, both $\thetasub$ and $\thetaipw$
are asymptotically linear. In addition, they are $n^{1/2}$-consistent
if the models are correctly specified. The bootstrap or an influence
function-based estimator may be used to construct asymptotically
correct confidence intervals. However, if the time-varying variables
are high-dimensional or there are too many time points such that
smoothing is necessary, the required consistency of $\hat\Q_t$ and
$\hat\r_t$ will hardly be achievable within parametric models. This
issue may be alleviated through the use of data-adaptive
estimators. Unfortunately, $n^{1/2}$-consistency of $\thetasub$ and
$\thetaipw$ will generally require that $\hat\Q_t$ and $\hat\r_t$ are
consistent in $L_2(\P)$-norm at parametric rate, which is generally
not possible when utilizing data-adaptive estimation of
high-dimensional regressions. Thus, the asymptotic distribution will
generally be unknown, making it difficult to construct confidence
intervals and hypothesis tests. In the following, we use the efficient
influence function to propose two estimators that are
$n^{1/2}$-consistent and efficient under weaker assumptions, requiring
only $n^{1/2}$-convergence of the second-order regression bias term
$\rem_0(\hat\eta)$.

\subsection{Targeted minimum loss-based estimator}\label{sec:tmle}

We start by presenting a generalization of the estimator we proposed for
the case of an LMTP in a single time point in
\cite{diaz2018stochastic}, which is also a generalization of the
estimator proposed by \cite{van2012targeted} for longitudinal dynamic
regimes. Compared to the estimator we proposed in \cite{Diaz12}, the
proposal in \cite{diaz2018stochastic} has the advantage that it does
not require a tilting model for the density $\g_t$, which may be
computationally intensive. In order to avoid imposing entropy
conditions on the initial estimators, we use sample splitting and
cross-fitting \citep{klaassen1987consistent,zheng2011cross,
  chernozhukov2018double}. Let ${\cal V}_1, \ldots, {\cal V}_J$ denote
a random partition of the index set $\{1, \ldots, n\}$ into $J$
prediction sets of approximately the same size. That is,
${\cal V}_j\subset \{1, \ldots, n\}$;
$\bigcup_{j=1}^J {\cal V}_j = \{1, \ldots, n\}$; and
${\cal V}_j\cap {\cal V}_{j'} = \emptyset$. In addition, for each $j$,
the associated training sample is given by
${\cal T}_j = \{1, \ldots, n\} \setminus {\cal V}_j$. We let
$\hat \eta_{j}$ denote the estimator of $\eta$ obtained by training
the corresponding prediction algorithm using only data in the sample
${\cal T}_j$. Further, we let $j(i)$ denote the index of the
validation set which contains observation $i$.

The targeted minimum loss-based estimator $\thetatmle$ is computed as
a substitution estimator that uses an estimate $\tilde\Q_{1,j(i)}$
carefully constructed to solve the cross-validated efficient influence
function estimating equation
$\Pn \{\D_1(\cdot, \tilde\eta_{j(\cdot)}) - \thetatmle\}= 0$. The
construction of $\tilde\Q_{1,j(i)}$ is motivated by the observation
that the efficient influence function of $\theta$ can be expressed as a
sum of terms of the form:
\[\left(\prod_{k=1}^t \r_k(a_k, h_k)\right)\{\Q_{t+1}(a_{t+1}^\d, h_{t+1}) -
  \Q_t(a_t, h_t)\},\] which take the form of score functions
$\omega(W)\{M - \E(M\mid W)\}$ for appropriately defined variables $M$
and $W$ and some weight function $\omega$. It is well known that if
$\E(M\mid W)$ is estimated within a weighted generalized linear model
with canonical link that includes an intercept, then the weighted MLE
estimate solves the score equation
$\sum_i\omega(W_i)\{M_i - \hat\E(M_i\mid W_i)\}=0$. TMLE uses this
observation to iteratively tilt preliminary estimates of $\Q_t$
towards a solution of the efficient influence function estimating
equation. The algorithm is defined as follows:

\begin{enumerate}[label=Step \arabic*., align=left, leftmargin=*]
\item Initialize $\tilde\eta =\hat\eta$ and
  $\tilde\Q_{\tau+1,j(i)}(A^\d_{\tau+1,i}, H_{\tau+1,i}) = Y_i$.
\item For $s=1,\ldots,\tau$, compute the weights
  \[\omega_{s,i} = \prod_{k=1}^s \hat\r_{k,j(i)}(A_{k,i}, H_{k,i})\]  
\item For $t=\tau,\ldots,1$:
  \begin{itemize}
  \item Fit the generalized linear tilting model
    \[\link \tilde\Q_t^\epsilon(A_{t,i},H_{t,i}) = \epsilon + \link
      \tilde\Q_{t,j(i)}(A_{t,1}, H_{t,i})\] where $\link(\cdot)$ is
    the canonical link. Here, $\link \tilde\Q_{t,j(i)}(a_t,h_t)$ is an
    offset variable (i.e., a variable with known parameter value equal
    to one). The parameter $\epsilon$ may be estimated by running a
    generalized linear model of the pseudo-outcome
    $\tilde \Q_{t+1,j(i)}(A^\d_{t+1,i}, H_{t+1,i})$ with only an
    intercept term, an offset term equal to
    $\link \tilde\Q_{t,j(i)}(A_{t,i},H_{t,i})$, and weights
    $\omega_{t,i}$, using all the data points in the sample. An
    outcome bounded in an interval $[a,b]$ may be analyzed with
    logistic regression (i.e., $\link=\logit$) by mapping it to an
    outcome $(0,1)$ through the transformation
    $(Y - a) / (b - a)(1-2\epsilon) + \epsilon$ for some small value
    $\epsilon>0$. This approach has robustness advantages compared to
    fitting a linear model as it guarantees that the predictions in
    the next step remain within the outcome space \citep[see
    ][]{Gruber2010t}.
  \item Let $\hat\epsilon$ denote the maximum likelihood estimate, and
    update the estimates as
    \begin{align*}
      \link \tilde\Q_{t, j(i)}^{\hat\epsilon}(A_{t,i},H_{t,i})&=
                                                                \hat\epsilon + \link\tilde\Q_{t, j(i)}(A_{t,i},H_{t,i})\\
      \link \tilde\Q_{t, j(i)}^{\hat\epsilon}(A_{t,i}^\d,H_{t,i})&=
                                                                   \hat\epsilon + \link\tilde\Q_{t, j(i)}(A_{t,i}^\d,H_{t,i}).
    \end{align*}
    The above procedure with canonical link guarantees the following
    score equation is solved: {\small
      \[\frac{1}{n}\sum_{i=1}^n \left(\prod_{k=1}^t \hat\r_{k,j(i)}(A_{k,i},
          H_{k,i})\right)\{\tilde \Q_{t+1,j(i)}(A^\d_{t+1,i}, H_{t+1,i}) -
        \tilde\Q_{t, j(i)}^{\hat\epsilon}(A_{t,i},H_{t,i})\}=0\]
    }
  \item Update $\tilde\Q_{t, j(i)}= \tilde\Q_{t, j(i)}^{\hat\epsilon}$, $t = t-1$, and iterate. 
  \end{itemize}
\item The TMLE is defined as
  \[\thetatmle=\frac{1}{n}\sum_{i=1}^n\tilde\Q_{1,j(i)}(A_{1,i}^\d,L_{1,i}).\]
\end{enumerate}
The iterative procedure and the score equation argument above
guarantee that
\[\frac{1}{n}\sum_{i=1}^n\sum_{t=1}^\tau\left(\prod_{k=1}^t \hat\r_{k,j(i)}(A_{k,i}, H_{k,i})\right)\{\tilde\Q_{t+1,j(i)}(A_{t+1,i}^\d, H_{t+1,i}) -
  \tilde\Q_{t,j(i)}(A_{t,i}, H_{t,i})\}=0,\] and thus that
$\Pn \{\D_1^{\tilde\eta_{j(\cdot)}} - \thetatmle\}= 0$. This fact is
crucial to prove the weak convergence result of $\thetatmle$ in
Theorem~\ref{theo:astmle}, which is useful to construct confidence
intervals and hypothesis tests.


\begin{theorem}[Weak convergence of TMLE]\label{theo:astmle}
  Assume the conditions of Theorem~\ref{theo:eif} hold. Assume
  that
  $\sum_{t=1}^\tau||\hat\r_t - \r_t||\, ||\tilde\Q_t - \Q_t|| =
  o_\P(n^{-1/2})$ and that
  $\P\{\r_t(A_t,H_t) < c\}=\P\{\hat\r_t(A_t,H_t) < c\}=1$ for some
  $c<\infty$. Then
  \[\sqrt{n}(\thetatmle - \theta) \rightsquigarrow
    N(0,\sigma^2),\] where
  $\sigma^2=\var\{\D_1(Z;\eta)\}$ is the non-parametric efficiency bound.
\end{theorem}

Note that $n^{1/2}$-consistency of the TMLE requires consistent
estimation of all nuisance parameters $(\r_t, \Q_t):t=1,\ldots,\tau$
at the rates stated in the theorem. The delta method implies these
rates would be trivially achieved if $\r_t$ and $\Q_t$ were estimated
within correctly pre-specified parametric models. The required rates
may also be achievable by many data-adaptive regression
algorithms. For example, these rates would be satisfied if
$||\hat\r_t - \r_t||=o_\P(n^{-1/4})$ and
$||\tilde\Q_t - \Q_t||=o_\P(n^{-1/4})$. For $\r_t$, the
$n^{-1/4}$-rate may be achievable by $\ell_1$ regularization
\citep{bickel2009simultaneous}, regression trees
\citep{wager2015adaptive}, neural networks \citep{chen1999improved},
or the highly adaptive lasso \citep{benkeser2016highly}. For $\Q_t$,
establishing the $n^{1/4}$-rate requires more careful analysis as the
outcome is fitted from data. Specifically, it is possible to bound the
regression error in terms of an error related to estimation of the
outcome plus an error purely due to regression estimation. Methods to
study this type of two-stage estimator can be found in
\cite{ai2003efficient,rubin2005general,foster2019orthogonal,kennedy2020optimal},
among others.

Beyond the $\sqrt{n}$-consistency implied by
Theorem~\ref{theo:astmle}, the TMLE is also multiply robust consistent
under $\tau+1$ configurations of consistency in estimation of the
nuisance parameters $\r_t$ and $\Q_t$.

\begin{lemma}[$\tau+1$ multiply robust consistency of
  TMLE]\label{theo:constmle}
  Assume the conditions of Theorem~\ref{theo:eif} hold. Assume that there is a time $k$ such that
  $||\tilde\Q_t - \Q_t|| = o_\P(1)$ for all $t > k$ and
  $||\hat\r_t - \r_t|| = o_\P(1)$ for all $t \leq k$. Then we have
  $\thetatmle =\theta + o_\P(1)$.
\end{lemma}

Lemma~\ref{theo:constmle} is a direct consequence of equation
(\ref{eq:exptmle}) in the supplementary materials together with the
expression for $\rem_1(\eta)$ in equation (\ref{eq:rem}).  Note that
although $\Q_t$ implicitly depends on $\g_{t+1},\ldots, \g_\tau$, it
is possible to construct estimators that achieve the conditions of the
lemma, as the parameterization (\ref{eq:defQ}) means $\Q_t$ and
$\g_{t+1},\ldots, \g_\tau$ are in fact variation independent, i.e.,
one can construct consistent estimators of $\Q_t$ without relying on
consistent estimators of $\g_{t+1},\ldots, \g_\tau$.

Inspection of $\rem_1(\eta)$ teaches us that a result stronger than
Lemma~\ref{theo:constmle} should be possible, i.e., consistent
estimation of $\theta$ should be achievable under a weaker sequential
doubly robust consistent (SDR) assumption that
$||\tilde\Q_t - \Q_t|| = o_\P(1) \,\vee\, ||\hat\r_t - \r_t|| =
o_\P(1)$ for all $t$. Note, however, that estimators of $\theta$ (such
as TMLE) that use the recursive definition in (\ref{eq:defQ}) to
estimate $\Q_t$ can only be expected to satisfy the assumption in the
Lemma~\ref{theo:constmle} (and not $2^\tau$-multiple robustness)
because $\Q_t$ cannot generally be consistently estimated using the
recursive definition (\ref{eq:defQ}) unless $\Q_s$ is consistently
estimated for all $s>t$. In the following we present a sequential
regression estimator that overcomes this limitation by using
expression (\ref{eq:fo}) instead of (\ref{eq:defQ}) to estimate
$\Q_t$. This fact motivates our use of the name \textit{sequentially
  doubly robust} since the estimator relies on the fact that each
$\Q_t$ can be consistently estimated even if $\Q_s$ is inconsistently
estimated for some $s>t$. This is in contrast to the more common name
\textit{\textit{$2^\tau$}-multiply robust}, which only applies to the
estimator of $\theta$.

\subsection{Sequential regression estimator using SDR unbiased
  transformations}\label{sec:sdr}

In this section we use the multiply robust unbiased transformation in
expression (\ref{eq:fo}) to obtain an estimate of $\Q_t$. This
estimator is an extension to LMTPs of estimators proposed by
\cite{luedtke2017sequential} and \cite{rotnitzky2017multiply} for
longitudinal dynamic regimes. We say that $\D_{t+1}$ is a multiply
robust unbiased transformation for $\Q_t$ due to the following
proposition, which is a straightforward consequence of
Lemma~\ref{theo:sdr}.
\begin{proposition}
  Let $\eta'$ be such that either $\Q'_s=\Q_s$ or $\r'_s=\r_s$ for all
  $s>t$. Then we have
  \[\E\big[\D_{t+1}(Z;\eta')\mid A_t=a_t,H_t=h_t\big] =\Q_t(a_t,h_t).\]
\end{proposition}
This lemma motivates the construction of the sequential regression
estimator by iteratively regressing an estimate of the data
transformation $\D_{t+1}(Z;\eta)$ on $(A_t,H_t)$, starting at
$\D_{\tau+1}(Z;\eta)=Y$. Similar ideas have been used by others to
obtain estimates of various causal inference parameters
\citep{buckley1979linear,rubin2007doubly,diaz2013targeted,kennedy2017non,luedtke2017sequential}. For
preliminary cross-fitted estimates
$\hat\r_{1,j(i)},\ldots,\hat \r_{\tau, j(i)}$, the estimator is
defined as follows:
\begin{enumerate}[label=Step \arabic*, align=left, leftmargin=*]
\item Initialize  $\D_{\tau+1}(Z_i;\underline{\check\eta}_{\tau,j(i)})= Y_i$ for $i=1,\ldots,n$.
\item For $t=\tau,\ldots,1$:
  \begin{itemize}
  \item Compute the pseudo-outcome
    $\check Y_{t+1,i} =
    \D_{t+1}(Z_i;\underline{\check\eta}_{t,j(i)})$ for all
    $i=1,\ldots,n$.
  \item For $j=1,\ldots,J$:
    \begin{itemize}
    \item Regress $\check Y_{t+1,i}$ on $(A_{t,i,}H_{t,i})$ using any
      regression technique and using only data points
      $i\in \mathcal T_{j}$.\label{step:2}
    \item Let $\check \Q_{t,j}$ denote the output, update
      $\underline{\check\eta}_{t, j} = (\hat\r_{t,j}, \check
      \Q_{t,j},\ldots,\hat\r_{\tau,j}, \check \Q_{\tau,j})$, and
      iterate.
    \end{itemize}
  \end{itemize}
\item Define the sequential estimator regression as
  \[\thetasr=\frac{1}{n}\sum_{i=1}^n\D_1(Z_i;\check\eta_{j(i)}).\]  
\end{enumerate}

To prove that the SDR estimator is sequentially doubly robust
consistent, it will be useful to have the following alternative
representation of $\rem_t(a_t,h_t;\eta)$. Define the data-dependent
parameter
\[\check\Q^\dagger(a_t, h_t) = \E\big[\D_{t+1}(Z;\underline{\check\eta}_t)\mid
  A_t=a_t,H_t=h_t\big],\] where the outer expectation is only with
respect of the distribution $\P$ of $Z$ (i.e., $\check\eta$ is fixed).
Equation (\ref{eq:fo}) yields
\begin{equation}
  \Q_t(a_t,h_t) = \check\Q_t^\dagger(a_t,h_t) +
  \rem_t(a_t, h_t;\check\eta).\label{eq:check}
\end{equation}
An induction argument together with (\ref{eq:rem}) yields the lemma below (proved in the
supplementary materials).
\begin{lemma}\label{lemma:remorder}
  Assume that $\P\{\r_t(A_t,H_t) < c\} =\P\{\hat\r_t(A_t,H_t) < c\}=1$ for some $c<\infty$. Then
  \begin{equation}
    \rem_0(\check\eta) = \sum_{t=1}^\tau O_\P\big(||\hat\r_t -
    \r_t||\,||\check\Q_t - \check\Q_t^\dagger||\big).\label{eq:rem2}
  \end{equation}
\end{lemma}
In comparison to (\ref{eq:rem}), the representation of the remainder
term in (\ref{eq:rem2}) avoids iterative definitions of the regression
error terms involved. Unlike $||\check\Q_t - \Q_t||$, which implicitly
depends on all $||\check\Q_s - \Q_s||:s > t$, the error term
$||\check\Q_t - \check\Q_t^\dagger||$ depends only on the consistency
of regression procedure used in \ref{step:2} to estimate the outer
expectation in expression (\ref{eq:check}). This representation is
thus more useful to establish sequential doubly robust consistency. In
particular, we have:
\begin{lemma}[$2^\tau$-multiply robust consistency of SDR
  estimator]\label{theo:conssdr} Assume the conditions of
  Theorem~\ref{theo:eif} hold. Assume
  that, for each time $t$, either $||\hat\r_t - \r_t|| = o_\P(1)$ or
  $||\check\Q_t - \check\Q^\dagger_t|| = o_\P(1)$ . Then we have
  $\thetasr =\theta + o_\P(1)$.
\end{lemma}

The sequential regression estimator also satisfies a weak convergence
result analogous to Theorem~\ref{theo:astmle}:

\begin{theorem}[Weak convergence of SDR
  estimator]\label{theo:asrem}

  Assume the conditions of Theorem~\ref{theo:eif} hold. Assume
  that
  $\sum_{t=1}^\tau||\hat\r_t - \r_t||\, ||\check\Q_t -
  \check\Q_t^\dagger|| =
  o_\P(n^{-1/2})$ and that
  $\P\{\r_t(A_t,H_t) < c\}=\P\{\hat\r_t(A_t,H_t) < c\}1$ for some
  $c<\infty$. Then
  \[\sqrt{n}(\thetasr - \theta) \rightsquigarrow
    N(0,\sigma^2),\] where
  $\sigma^2=\var\{\D_1(Z;\eta)\}$ is the non-parametric efficiency bound.
\end{theorem}

Like its TMLE counterpart, the above theorem may be used to construct
asymptotically valid confidence intervals and hypothesis tests. For
example, the standard error may be estimated as the empirical variance
of $\D_1(Z;\check\eta_{j(i)})$, and this standard error may
be used to compute Wald-type confidence
intervals. 

We note that the rates required for $n^{1/2}$-consistency in
Theorems~\ref{theo:astmle} and \ref{theo:asrem} are the same for the
TMLE and SDR, and in this sense the SDR estimator does not seem to
confer asymptotic advantages with respect to the TMLE. Furthermore,
the TMLE and SDR estimators proposed in this paper do not allow for
the construction of multiply robust confidence regions or p-values.
Much of the recent literature focused on solving this problem proposes
the construction of estimators that remain $n^{1/2}$-consistent under
multiply robust consistency assumptions on the nuisance
estimators. The interested reader is referred to
\cite{van2014targeted,farrell2015robust,benkeser2016doubly,
  diaz2017doubly,diaz2019statistical,smucler2019unifying}, and
references therein as examples of this literature. The SDR and TMLE
estimators proposed here could possibly be adapted to satisfy this
property through extension of the methods of
\cite{diaz2019statistical} to LMTPs.

Lemmas~\ref{theo:constmle} and \ref{theo:conssdr} only state the
conditions for consistency and do not provide the convergence rate of
the TMLE and SDR under misspecification of the nuisance
parameters. For example, for $\tau=2$ with $\Q_2$ inconsistently
estimated we know the TMLE will always be inconsistent and the SDR
estimator may be consistent if the other conditions of
Lemma~\ref{theo:conssdr} hold. However, the rate of consistency of the
SDR in this case is unknown. We conjecture that the additional
robustness of the SDR conferred by Lemma~\ref{theo:conssdr} endows
this estimator with better finite sample behavior compared to the
TMLE. We explore this conjecture in our simulation studies in
\S\ref{sec:sim} of the supplementary materials.

An important possible drawback of the sequential regression estimator
is that the pseudo-outcome $\check Y_{t,i}$ may be outside the bounds
of the original outcome, which may in turn yield a parameter estimate
$\thetasr$ out of bounds of the parameter space. This may be
especially problematic if the intervention $\d$ is allowed to yield
near violations of the positivity assumption, in which case the
density ratio $\r_t$ may be highly variable. While this can be
remedied using truncation, an extension of the sequential regression
infinite-dimensional TMLE strategy proposed by
\cite{luedtke2017sequential} for the case of a static intervention may
offer a more principled solution.

\subsection{Density ratio estimation via
  classification}\label{sec:densratio}

The estimators proposed in the previous sections require a preliminary
estimator of the density ratio $\r_t$. A possible strategy to
estimate this density ratio is to obtain estimates of the density
$\g_t$ and plug them into the definition of $\g_t^\d$ to compute the
ratio. This approach may fail for several reasons, the most important
being that data-adaptive estimators for high-dimensional conditional
densities are scarce in the machine and statistical learning
literature. In this section we propose a different approach, in which
we recast the problem of estimating the density ratio $\r_t$ in terms
of a classification problem in an augmented dataset that contains $2n$
observations \citep{qin1998inferences, cheng2004semiparametric}.

For a fixed time point, consider an augmented dataset of size $2n$ in
which we have duplicated each observation. In this augmented dataset,
one of the duplicated records gets assigned the actually observed
exposure, $A_t$, and the other one gets assigned exposure under the
intervention, $A_t^\d$. We also introduce an indicator variable
$\Lambda$, which is equal to one if the duplicated observation
corresponds to the treatment under intervention, and zero
otherwise. The augmented dataset can be represented as follows:
$(H_{\lambda,i,t}, A_{\lambda,i,t} ,
\Lambda_{\lambda,i}:\lambda=0,1;i=1,\ldots,n)$, where
$\Lambda_{\lambda,i}=\lambda$ indexes the duplicates,
$H_{\lambda,i,t} = H_{i,t}$ is the history variable and is equal for
both duplicated records, and
$A_{\lambda,i,t} = \lambda\times A_{i,t}^\d + (1-\lambda)\times
A_{i,t}$ is the natural exposure level if $\lambda = 0$, and the
intervened exposure level if $\lambda = 1$.

Denote the probability distribution of $(H_t, A_t, \Lambda)$ in the
augmented dataset by $\P^\lambda$, and the corresponding density by
$\p^\lambda$. Define the following parameter of $\P^\lambda$:
\[\uu^\lambda_t(a_t, h_t) = \P^\lambda(\Lambda = 1\mid A_t=a_t,
  H_t=h_t).\]
Then, we have the following relation between the density
ratio $\r_t$ and the distribution $\P^\lambda$:
\[\r_t(a_t, h_t) = \frac{\p^\lambda(a_t, h_t \mid \Lambda =
    1)}{\p^\lambda(a_t, h_t \mid \Lambda =
    0)}=\frac{\P^\lambda(\Lambda = 1\mid A_t=a_t,
    H_t=h_t)}{\P^\lambda(\Lambda = 0\mid A_t=a_t,
    H_t=h_t)}=\frac{\uu^\lambda_t(a_t,h_t)}{1-\uu^\lambda_t(a_t,h_t)},\]
where the first equality follows by definition of $\r_t$ and the
definition of conditional density, the second follows by Bayes rule
and the observation that
$1/2=\P^\lambda(\Lambda = 1)=\P^\lambda(\Lambda = 0)$, and the last
one by definition. Thus, estimation of the density ratio may be
carried out by estimating $\uu^\lambda$ in the augmented dataset via
any classification method available in the machine and statistical
learning literature (e.g., Super Learning). In order to preserve
properties of the estimator such as the asymptotic normality presented
in Theorems~\ref{theo:astmle} and \ref{theo:asrem}, cross-validation
and cross-fitting should be performed as follows. Let
${\cal V}_1, \ldots, {\cal V}_J$ denote a random partition of the
index set $\{1, \ldots, n\}$ as in \S\ref{sec:tmle}, and let
${\cal T}_j = \{1, \ldots, n\} \setminus {\cal V}_j$. The estimator
$\hat\r_{t,j(i)}(A_{i,t},H_{i,t})$ may be computed as
$\hat
\uu_{t,j(i)}^\lambda(A_{i,t},H_{i,t})/\{1-\hat\uu_{t,j(i)}^\lambda(A_{i,t},
H_{i,t})\}$ where $\uu_{t,j(i)}^\lambda$ is the result of training a
classification algorithm using data
$(A_{\lambda,i,t} , H_{\lambda,i,t},
\Lambda_{\lambda,i}:\lambda=0,1;i\in \mathcal T_{j(i)})$. This
cross-validation scheme is necessary because the rows in the augmented
dataset are not i.i.d. Performing cross-validation on the index set
$\{i:1,\ldots,n\}$ rather than $\{\lambda=0,1;i=1,\ldots,n \}$ ensures
that certain independencies required for the proofs of
Theorems~\ref{theo:astmle} and \ref{theo:asrem} remain true.
\section{Illustrative application}

To illustrate the proposed methods, we estimate the effect of an
intervention on patients with acute respiratory failure on 14-day
survival among 10,044 intubated ICU patients. The data used in this
illustration is the Weill Cornell Critical carE Database for Advanced
Research (WC-CEDAR), a comprehensive data repository containing
demographic, laboratory, procedure, diagnosis, medication,
microbiology, and flow sheet data documented as part of standard
care. I

\begin{figure}[H]
  \centering
  \includegraphics[scale = 0.8]{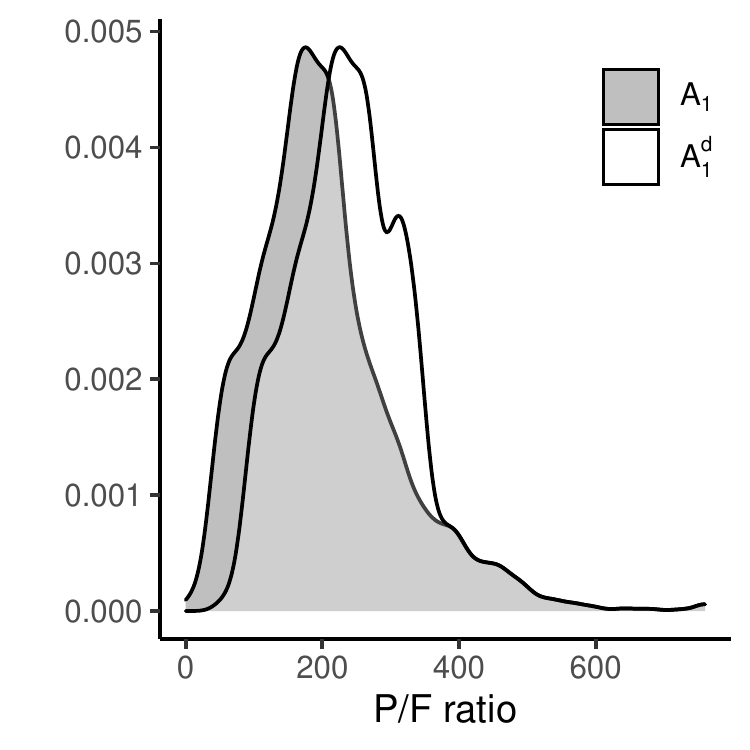}
  \caption{Pre- and post-intervention distribution of the P/F ratio at
    day 1 in our illustrative application.}\label{fig:example}
\end{figure}

The study cohort includes patients who are invasively mechanically
ventilated. Study time begins at their time of intubation. The
exposure of interest is the worst daily arterial partial pressure of
oxygen to fraction of inspired oxygen (PaO2 to Fio2, or P/F)
ratio. The P/F ratio is a continuous measure of hypoxemia and is used
to quantify the severity of acute respiratory failure. For example,
the Berlin definition of Acute Respiratory Distress Syndrome (ARDS)
uses a P/F ratio of $<$ 100 for its classification of severe ARDS, 100
$\leq$ P/F ratio $<$ 200 for moderate ARDS, and 200 $\leq$ P/F ratio
$<$ 300 for mild.  There are several ways physicians can intervene on
an individual patient's P/F ratio, but the most direct modifications
are through supplemental oxygen, e.g. invasive and non-invasive forms
of mechanical ventilation. When a patient is mechanically ventilated,
various changes to device parameters can significantly affect the
patient's P/F ratio. Among other interventions to increase patients'
P/F ratios are neuromuscular blockade medications, physical techniques
such as prone positioning, and treating the underlying condition
e.g. administering loop diuretic, antibiotics or steroids. Some of
these interventions have negative effects such as hypotension and
renal failure. Our goal is to estimate the overall effect on survival
from the time of invasive mechanical ventilation of an intervention
that would increase the P/F ratio by 50 among patients with clinically
defined acute respiratory failure (P/F ratio $<$ 300). An increase of
50 is both clinically feasible and clinically
meaningful. Figure~\ref{fig:example} shows an example of the pre- and
post-intervention distribution of the P/F ratio for day 1.

Baseline confounders include age, sex, race, number of Elixhauser
comorbidities, and pneumonia status. Time-dependent confounders
include daily mechanical ventilation status (invasive or non-invasive)
and the Sequential Organ Failure Assessment (SOFA) score with the
pulmonary component removed. 
Patients are censored at their day of hospital discharge, as vital
status was unknown after this point.

We estimate the effect of the LMTP using substitution, IPW, TMLE, and
SDR estimation. The functions $\r_t$ and $\Q_t$ are estimated using
the Super Learner \citep{vanderLaan&Polley&Hubbard07}. We included
multivariate adaptive regression splines (MARS), extreme gradient
boosting, logistic regression with $\ell_1$ regularization (LASSO),
and simple logistic regression as our candidate learners. Importantly,
the estimators of $\r_t$ and $\Q_t$ condition on the complete history
of all variables and do not make Markov assumptions. The average
weights given to each learner at each of the 14 time points for $\r$
and $\Q$ are shown in Figure~\ref{fig:weights} in the supplementary
materials.
Using SDR to adjust for right censoring, we estimated the 14-day
survival under no intervention on P/F ratios to be 88.3\% (95\% CI
87.5, 89.0\%). The estimate under an increase of 50 among hypoxic
patients is estimated as 89.8\% (95\% CI 88.0, 91.5\%). This
intervention would increase 14-day survival by 1.5\% (95\% CI -0.1,
3\%). All results, including those using the TMLE, substitution, and
IPW estimators, can be found in Table \ref{tab:respf}.

\begin{table}[ht]
  \centering
  \captionsetup{font=scriptsize}
  \caption{Estimated 28-day ICU mortality under the observed P/F
    ratios and a treatment policy that increases P/F ratio by 50 units
    when below 300.}
  \resizebox{\linewidth}{!}{%
    \begin{tabular}{lcccccccccc}
      \toprule
              & \multicolumn{3}{c}{No shift} & \multicolumn{3}{c}{Trt. Policy} & \multicolumn{4}{c}{Trt. Effect}                                                                    \\
      \cmidrule(l{3pt}r{3pt}){2-4} \cmidrule(l{3pt}r{3pt}){5-7}
      \cmidrule(l{3pt}r{3pt}){8-11}
      
       Estimator & Estimate & SE    & 95\% CI        & Estimate & SE    & 95\% CI        & Estimate & SE    & 95\% CI         & P     \\
      \midrule
       IPW  & 0.754 & -     & (-, -)         & 0.978 & -     & (-, -)         & 0.223 & -     & (-, -)          & -     \\
       Sub. & 0.877 & -     & (-,  -)        & 0.891 & -     & (-, -)         & 0.014 & -     & (-, -)          & -     \\
       TMLE & 0.888 & 0.004 & (0.880, 0.895) & 0.896 & 0.009 & (0.879, 0.914) & 0.009 & 0.008 & (-0.006, 0.024) & 0.244 \\
       SDR  & 0.883 & 0.004 & (0.875, 0.890) & 0.898 & 0.009 & (0.880, 0.915) & 0.015 & 0.008 & (-0.001, 0.030) & 0.053 \\
      \bottomrule
    \end{tabular}%
  }
  \label{tab:respf}
\end{table}

While the substitution estimator as well as the TMLE and SDR
estimators produced comparable results, the IPW estimates an increase
of 22\% in survival. The disagreement between the IPW and other
estimators is due to the high variability of the density ratios, which
make the IPW estimator highly variable. For example, the weights
$\prod_{t=1}^\tau\hat\r_t(A_t,H_t)$ have a coefficient of variation of
5.5\%. Here we note that this instability of the IPW does not mean
that there are positivity violations in our application (the maximum
value of $\prod_{t=1}^\tau\hat\r_t(A_t,H_t)$ is 97.7), but rather that
the weights are highly variable and so is the IPW estimator. The
distribution of the weights for the IPW estimator is presented in
Figure~\ref{fig:ipw} in the supplementary materials.


\bibliographystyle{plainnat}
\bibliography{refs}

\end{document}


\maketitle
\section{Simulation study}\label{sec:sim}

We present a numerical study using Monte Carlo simulation to
illustrate the properties of the proposed estimators. For each sample
size $n\in\{200,800,1800,3200,5000\}$ and $\tau=4$, we generated 1000
datasets from the following data generating mechanism:
\begin{align*}
  L_1 \sim & \mathtt{Cat}(0.5, 0.25, 0.25)\\
  A_1\mid L_1 \sim & \mathtt{Binomial}\{5, \one(L_1>1)0.5
                     +\one(L_1>2)0.1\}\\
  L_t\mid (\bar A_{t-1},  \bar L_{t-1})\sim &
                                              \mathtt{Bernoulli}\{\expit(-0.3L_{t-1}
                                              + 0.5A_{t-1})\}\text{ for }
                                              t \in\{2,3,4\}\\
  A_t\mid (\bar A_{t-1},  \bar L_t)\sim &
                                          \begin{cases}\mathtt{Binomial}\{5,
                                            \expit(-2+1/(1 + 2 L_t +
                                            A_{t-1}))\}\text{ for }
                                            t \in\{2,3\}\\
                                            \mathtt{Binomial}\{5,
                                            \expit(1 + L_t - 3A_{t-1})\}\text{ for }
                                            t = 4
                                          \end{cases}\\
  Y\mid (\bar A_4, \bar L_4)\sim & \mathtt{Bernoulli}\{\expit(-2 + 1/(1
                                   - 1.2A_4 - 0.3L_4))\},
\end{align*}
where $\mathtt{Cat}(p_1,p_2,p_3)$ is a categorical distribution taking
values 1,2,3 with probabilities $p_1,p_2,p_3$, respectively;
$\mathtt{Bernoulli}\{p\}$ is a Bernoulli distribution with probability
$p$; and $\mathtt{Binomial}\{k,p\}$ is a Binomial distribution with
$k$ trials and probability $p$. For illustration, we assess the
performance of the estimators for estimating the effect of the
following LMTP:
\begin{equation*}
  \d(a_t,h_t)=
  \begin{cases}
    a_t - 1 & \text{if } a_t \geq 1 \\
    a_t & \text{if } a_t < 1,
  \end{cases}
\end{equation*}
where we have not intervened on observations whose natural exposure
value is $a_t=0$, since they would have an implausible exposure value
of $\d(a_t,h_t)=-1$ under intervention. 

\begin{figure}[H]
  \centering
  \includegraphics[scale = 0.08]{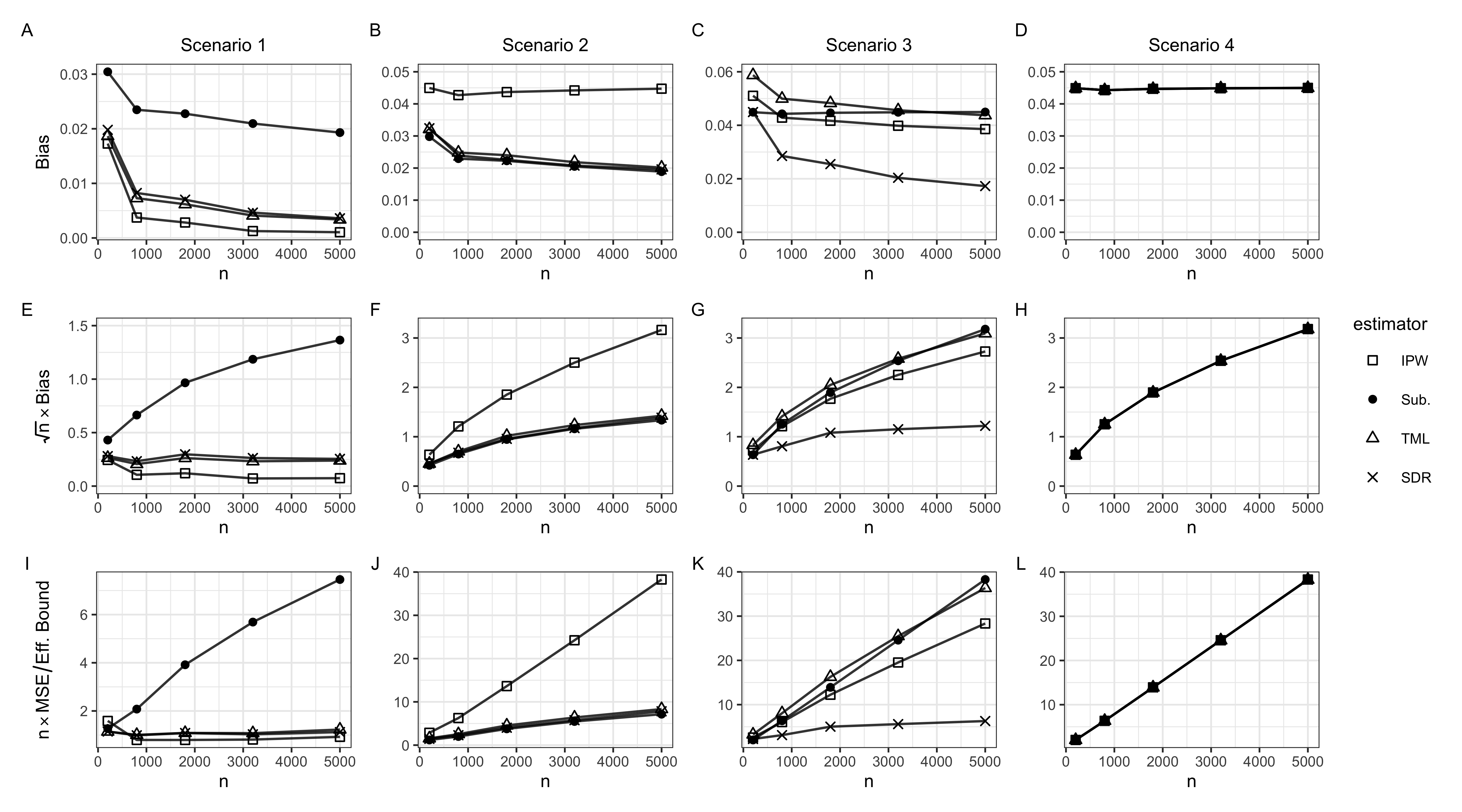}
  \caption{Performance of several estimators of the effect of a
    longitudinal modified treatment policy.}\label{fig:bias}
\end{figure}

\begin{figure}[H]
  \centering
  \includegraphics[scale = 0.08]{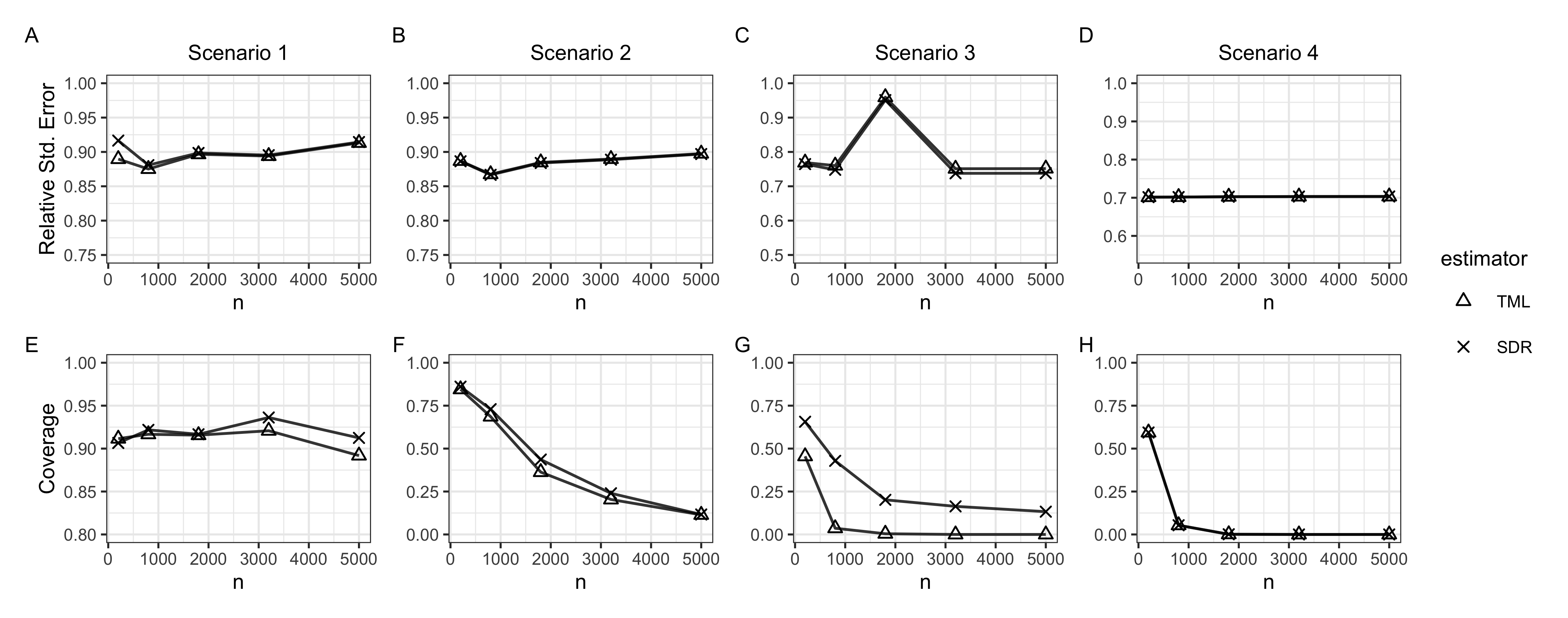}
  \caption{Performance of the standard error estimator and Wald-type
    95\% confidence interval for the TMLE and SDR
    estimators. Relative standard error is the mean of the
    S.E. compared to the efficiency bound.}\label{fig:coverage}
\end{figure}

We tested the four estimators proposed in four scenarios: (1) All
nuisance parameters are consistently estimated. (2) The estimators of
$\Q_t$ are consistent for $t>2$ but inconsistent otherwise, and the
estimators for $\r_t$ are consistent for $t\leq 2$ but inconsistent
otherwise. (3) The estimators of $\Q_t$ are consistent for $t<4$ but
inconsistent for $t=4$, and the estimators for $\r_t$ are inconsistent
for $t<4$ but consistent for $t=4$. (4) All estimators are
inconsistent. Consistency is achieved by fitting the highly adaptive
lasso (HAL), while inconsistent estimators ignore all covariates and fit an
intercept-only model. While this type of inconsistency is unlikely in
practice, it serves the purpose of illustrating the properties of the
proposed estimators. The HAL estimator is fitted agnostic of the
Markov property of the data generating mechanism. Note that, even
though all the variables are categorical, a non-parametric MLE
estimator of $\Q_\tau$ is infeasible since it would involve
$6^4\times 2^3\times 3= 31114$ empirical means. We compare the results
in terms of several metrics: bias, bias scaled by $n^{1/2}$, and MSE
relative to the efficiency bound scaled by $n$. These results are
presented in Figure~\ref{fig:bias}. In addition, we also assess the
performance of the influence function based standard error estimator
$\hat\sigma/\sqrt{n}$ in terms of the estimate $\hat\sigma$ relative
to the squared root of the efficiency bound $\sigma$ and coverage of a
Wald-type 95\% confidence interval that uses $\hat\sigma$. These
results are presented in Figure~\ref{fig:coverage}.

The results in Figure~\ref{fig:bias} are largely what should be
expected according to theory.  When all estimators are consistent, the
$n^{1/2}$-bias of the TMLE and SDR converges to zero (panel E), and
the estimators achieve the non-parametric efficiency bound (panel
I). Both the TMLE and SDR are consistent when the conditions of both
Lemma~\ref{theo:constmle} and Lemma~\ref{theo:conssdr} are satisfied
(panel B). However when only the conditions of
Lemma~\ref{theo:conssdr} but not of Lemma~\ref{theo:constmle} are
satisfied (panel C), only the SDR estimator seems consistent. This
illustrates the additional benefits of $2^\tau$-multiple robustness,
compared to $\tau+1$-multiple robustness. The substituton estimator
seems consistent when all estimators of $\Q_t$ are consistent (panel
A), but its bias seems not to converge at $n^{1/2}$-rate. When all
estimators of $\r_t$ are consistent, the IPW seems consistent (panel
A) and $n^{1/2}$ consistent (panel I), though the latter should not be
expected in general. The IPW estimator is inconsistent if at least one
of the estimators for $\r_t$ is inconsistent (panels B, C, D).
All estimators are inconsistent if all models are misspecified
(panel D).
The results of Figure~\ref{fig:coverage} also highlight a few
important characteristics of the standard error estimators in this
simulation. First, with all models correctly specified, the empirical
standard deviation of the EIF seems to be an underestimate of the
standard error at all sample sizes we consider. This yields slightly
below nominal coverage for both the TMLE and SDR estimators at all
sample sizes evaluated. In scenarios 2, 3, and 4, the $n^{1/2}$-bias
of both estimators yields confidence intervals with zero asymptotic
coverage, with worse performance for the TMLE in scenario 3.

\section{Identification (Theorem \ref{theo:iden})}
\begin{proof}
  The proof of part (i) follows directly from Lemma 1 in the Appendix
  of \cite{kennedy2018nonparametric}. Specifically, let
  \begin{equation}
    \g_t^\d(a_t\mid h_t)=\int\int_{a'_t:\d(a'_t,h_t,\epsilon_t)=a_t}\g(a_t'\mid
    h_t)\dd\nu(a_t')\dd\P(\epsilon_t),\label{eq:defgtd}
    \end{equation} and let $Q_t^\d$ denote a draw from
  $\g_t^\d(\cdot \mid H_t)$.  \cite{kennedy2018nonparametric} shows
  that $\E[Y(\bar Q^\d)]$ is identified as
  \[\E[Y(\bar Q^\d)]=\int_{\mathcal {\bar A}_\tau, \mathcal {\bar
        L}_\tau}\E[Y\mid A_\tau = a_\tau, H_\tau =
    h_\tau]\prod_{k=1}^\tau\g_k^\d(a_k\mid h_k) \dd
    \nu(a_k)\dd\P(l_k\mid a_{k-1},h_{k-1}),\] under Assumptions
  \ref{ass:support} and \ref{ass:exch1}. The result of the theorem
  follows after plugging in (\ref{eq:defgtd}) in the above expression.
  
  The proof of part (ii) is as follows.  Let $A_{t+1}(\bar
  a_t)$ and $L_{t+1}(\bar
  a_t)$ denote the counterfactual variables at time
  $t+1$ in a hypothetical world where $\bar A_t = \bar
  a_t$. Fix a time point $t$. For $s=1$, let
  \[A_{t+s}^{\d, \dagger}(\bar a_t, \bar l_t) = \d\{ \bar a_t, A_{t+s}(\bar
    a_t), \bar l_t, L_{t+s}(\bar a_t)\},\] and for $s>1$ define
  $A_{t+s}^{\d, \dagger}$ recursively as
  \begin{align*}
    A_{t+s}^{\d, \dagger}(\bar a_t, \bar l_t) = \d\{
    &\bar a_t,  A_{t+1}(\bar
      a_t), A_{t+2}(\bar a_t, A_{t+1}^{\d, \dagger}(\bar a_t)), \ldots,
      A_{t+s}(\bar a_t, A_{t+1}^{\d, \dagger}(\bar a_t), \ldots,
      A_{t+s-1}^{\d, \dagger}(\bar a_t)), \\
    &\bar l_t,  L_{t+1}(\bar
      a_t), L_{t+2}(\bar a_t, A_{t+1}^{\d, \dagger}(\bar a_t)), \ldots,
      L_{t+s}(\bar a_t, A_{t+1}^{\d, \dagger}(\bar a_t), \ldots,
      A_{t+s-1}^{\d, \dagger}(\bar a_t))\}.
  \end{align*}
  Likewise define
  \[L_{t+s}^\dagger(\bar a_t) = L_{t+s}(\bar a_t, A_{t+1}^{\d, \dagger}(\bar a_t), \ldots,
    A_{t+s-1}^{\d, \dagger}(\bar a_t)),\]
  and define the random variable
  \[    Z_t(\bar a_t, \bar l_t)= f_Y(\bar a_t, \underline A_{t+1}^{\d, \dagger}(\bar a_t, \bar l_t)
    ,
    \bar l_t,
    \underline L_{t+1}^\dagger(\bar a_t), 
    U_Y).
  \]
  Recall that for fixed values $\bar a_t$,
  $\bar l_t$, we denote
  $a_t^\d=\d(a_t, h^\d_t)$, where
  $h^\d_t=(\bar a_{t-1}^\d, \bar l_t)$. We
  have
  \begin{align}
    E[Y(\bar A^\d)] & =\int_{\mathcal A_1, \mathcal L_1} \E[Z_1(a_1^\d, l_1)\mid A_1=a_1, L_1=l_1]\dd
               \P(a_1,l_1)\notag\\
             & =\int_{\mathcal A_1, \mathcal L_1} \E[Z_1(a_1^\d, l_1)\mid A_1=a_1^\d, L_1=l_1]\dd
               \P(a_1,l_1)\notag\\
             & =\int_{\mathcal {\bar A}_2, \mathcal {\bar L}_2}
               \E[Z_1(a_1^\d, l_1)\mid A_2=a_2,
               L_2=l_2,A_1=a_1^\d,L_1=l_1]\dd \P(a_2,l_2\mid a_1^\d,
               l_1)\dd \P(a_1,l_1)\notag\\
             & =\int_{\mathcal {\bar A}_2, \mathcal {\bar L}_2}
               \E[Z_2(\bar a_2^\d, \bar l_2)\mid A_2=a_2,
               L_2=l_2,A_1=a_1^\d,L_1=l_1]\dd \P(a_2,l_2\mid a_1^\d,
               l_1)\dd \P(a_1,l_1)\notag\\
             & =\int_{\mathcal {\bar A}_2, \mathcal {\bar L}_2}
               \E[Z_2(\bar a_2^\d, \bar l_2)\mid A_2=a_2^\d,
               H_2=h_2^\d]\dd \P(a_2,l_2\mid a_1^\d,
               h_1^\d)\dd \P(a_1,l_1)\notag\\
             &\,\,\, \vdots\notag\\
             & =\int_{\mathcal {\bar A}_\tau, \mathcal {\bar L}_\tau}
               \E[Z_\tau(\bar a_{\tau}^\d, \bar l_{\tau})\mid A_\tau=a_\tau^\d,
               H_\tau=h_\tau^\d] \prod_{k=1}^\tau\dd\P(a_k,l_k\mid
               a_{k-1}^\d, h_{k-1}^\d)\notag\\
             & =\int_{\mathcal {\bar A}_\tau, \mathcal {\bar L}_\tau} \E[Y\mid A_\tau=a_\tau^\d,
               H_\tau=h_\tau^\d] \prod_{k=1}^\tau\dd\P(a_k,l_k\mid
               a_{k-1}^\d, h_{k-1}^\d)\label{eq:eq1}
  \end{align}
  Where the first and third equalities follow by law of iterated
  expectation and the definition of $Y(\d)$, the second and fifth
  follow from Lemma~\ref{lemma:indep} below and Assumptions~\ref{ass:exch} and
  \ref{ass:support}, the fourth one because
  $Z_t(\bar a_t^\d, \bar l_t)=Z_{t+1}(\bar a_{t+1}^\d, \bar
  l_{t+1})$ in the event $A_{t+1}=a_{t+1}, H_{t+1}=h_{t+1}^\d$, and the last
  one by definition. The definition in the theorem follows from
  recursive evaluation of the above integrals applying
  \begin{align*}
    \int_{\mathcal A_t, \mathcal L_t}\Q_t( a_t^\d,
    h_t^\d) \dd\P(a_t,l_t\mid a_{t-1}^\d, h_{t-1}^\d) &=\E[\Q(A_t^\d,
                                                       H_t)\mid A_{t-1} = a_{t-1}^\d, H_{t-1} = h_{t-1}^\d]\\
                                                     &=\Q_{t-1}(a_{t-1}^\d,
                                                       h_{t-1}^\d)
  \end{align*}
  starting with $t=\tau$.
\end{proof}

\begin{lemma}\label{lemma:indep}
  Under Assumption~\ref{ass:exch}, we have $Z_t(\bar a_{t-1}, \bar l_{t-1})\indep
  A_t\mid H_t$.
\end{lemma}
\begin{proof}
  This lemma follows trivially after noticing that under the assumed
  NPSEM, $Z_t(\bar a_{t-1}, \bar l_{t-1})$ is a deterministic function
  of $(\underline U_{L,t+1}, \underline U_{A,t+1})$.
\end{proof}
\section{Efficient influence functions (Theorem \ref{theo:eif})}
\begin{proof}
  In this proof we will use $\Theta(\P)$ to denote a parameter
  as a functional that maps the distribution $\P$ in the model to a
  real number. We will prove the result assuming $\d$ is smooth and
  invertible in all the range of $A$, the proof under
  Assumption~\ref{ass:inv} follows by partitioning the range of $A$ as
  in the assumption. A function $\D_1(Z;\eta)$ is the EIF of a
  parameter functional $\Theta(P)$ if it satisfies
  \begin{equation}
    \frac{\dd}{\dd  \epsilon}\Theta(P_\epsilon)\bigg|_{\epsilon=0} =
    \E[\D_1(Z;\eta) s(Z)],\label{eq:defeif}
  \end{equation}
  where $\P_\epsilon$ is a smooth parametric submodel that locally
  covers the non-parametric model, with score
  \[s(Z)=\left(\frac{\dd \log \P_\epsilon}{\dd
        \epsilon}\right)\bigg|_{\epsilon = 0}\] such that
  $\P_{\epsilon=0}=\P$. We start by conjecturing an EIF, and then
  prove it is in fact the EIF using the above definition.
  
  To conjecture an EIF we find the EIF in a model where we assume that
  the measure $\nu$ is discrete so that integrals can be written as
  sums. Using Lemma~\ref{lemma:altrep} below, we can express the
  non-parametric MLE of $\theta$ as
  \begin{equation}
    \Theta(\Pn)=\sum_{\bar a_\tau, \bar l_{\tau+1}}l_{\tau+1}\Pn
    f_{l_{\tau+1},a_\tau,h_\tau}\prod_{k=1}^\tau\frac{\Pn f_{\d_k,
        h_k}}{\Pn f_{a_k, h_k}}\label{nonpest},
  \end{equation}
  where we remind the reader of the notation $\P f =\int f
  \dd\P$. Here
  $f_{l_{\tau+1},a_\tau,h_\tau}(Z)=\one(L_{\tau+1} = l_{\tau+1},A_\tau
  = a_\tau,H_\tau = h_\tau)$, and $\one(\cdot)$ denotes the indicator
  function. We also define
  $f_{\d_k, h_k}(Z) = \one\{\d(A_k, h_k)=a_k, H_k=h_k\}$ and
  $f_{a_k, h_k}(Z) = \one\{A_k=a_k, H_k=h_k\}$.  In the above display
  we used the assumption that $\d$ does not depend on $\P$, and
  do not need to be estimated.

  We will use the fact that the efficient influence function in a
  non-parametric model corresponds with the influence curve of the
  NPMLE. This is true because the influence curve of any regular
  estimator is also a gradient, and a non-parametric model has only
  one gradient. The Delta method \cite[see, e.g., Appendix 18
  of][]{vanderLaanRose11} shows that if $\hat \Theta(\Pn)$ is a
  substitution estimator such that $\theta=\hat \Theta(\P)$, and
  $\hat \Theta(\Pn)$ can be written as
  $\hat \Theta^*(\Pn f:f\in\mathcal{F})$ for some class of functions
  $\mathcal{F}$ and some mapping $\Theta^*$, then the influence function of
  $\hat \Theta(\Pn)$ is equal to
  \[\dr_\P(Z)=\sum_{f\in\mathcal{F}}\frac{\dd\hat \Theta^*(\P)}{\dd\P f}\{f(O)-\P f\}.\]

  Applying this result to (\ref{nonpest}) with
  $\mathcal{F}=\{f_{l_{\tau+1},a_\tau,h_\tau}, f_{\d_s, h_s},
  f_{a_s, h_s}: l_{\tau+1}, a_\tau, h_\tau; s=1,\ldots,\tau\}$ and
  rearranging terms gives $\dr_P(Z) = \D_1(Z;\eta)$. In particular we have
  \begin{align*}
    \frac{\dd \Theta(\P)}{\dd f_{l_{\tau+1},a_\tau,h_\tau}} &=
    l_{\tau+1}\prod_{k=1}^\tau\frac{\P f_{\d_k, h_k}}{\P f_{a_k,
    h_k}}\\
    &=    l_{\tau+1}\prod_{k=1}^\tau\r(a_k, h_k),
  \end{align*}
and 
\begin{align*}
  \frac{\dd \Theta(\P)}{\dd \P f_{\d_s,h_s}}
  &= \sum_{l_{\tau+1},
    \underline a_{s+1},
    \underline
    h_{s+1}}l_{\tau+1}\frac{\P
    f_{l_{\tau+1},a_\tau,h_\tau}}{\P
    f_{\d_s, h_s}}\prod_{k=1}^\tau\frac{\P f_{\d_k, h_k}}{\P f_{a_k,
    h_k}}\\
  &= \sum_{\underline a_{s+1},
    \underline
    h_{s+1}}\Q_\tau(a_\tau, h_\tau)\frac{\P
    f_{a_\tau,h_\tau}}{\P
    f_{\d_s,
    h_s}}\prod_{k=1}^\tau\r(a_k,
    h_k)\\
  &= \sum_{\underline a_{s+1},
    \underline
    h_{s+1}}\Q_\tau(a_\tau, h_\tau)\frac{\P
    f_{a_\tau,h_\tau}}{\P
    f_{\d_s,
    h_s}}\prod_{k=1}^\tau\r(a_k,
    h_k)\\
  &=  \left(\sum_{\underline a_{s+1},
    \underline
    h_{s+1}}\Q_\tau(a_\tau, h_\tau)\frac{\P
    f_{a_\tau,h_\tau}}{\P
    f_{\d_s,
    h_s}}\prod_{k=s+1}^{\tau}\r(a_k,
    h_k)\right)\left(\prod_{k=1}^s\r(a_k,
    h_k)\right)\\
  &=  \left(\sum_{\underline a_{s+1},
    \underline
    h_{s+1}}\Q_\tau(a_\tau, h_\tau)\frac{\P
    f_{a_\tau,h_\tau}}{\P
    f_{a_s, h_s}}\prod_{k=s+1}^{\tau}\r(a_k,
    h_k)\right) \frac{\P
    f_{a_s,h_s}}{\P f_{\d_s,
    h_s}}\left(\prod_{k=1}^s\r(a_k, h_k)\right)\\
  &= \Q_s(a_s, h_s)\prod_{k=1}^{s-1}\r(a_k,
    h_k), 
\end{align*}
as well as
\begin{align*}
  \frac{\dd \Theta(\P)}{\dd \P f_{a_s,h_s}} &= -\sum_{l_{\tau+1},
                                            \underline a_{s+1},
                                            \underline
                                            h_{s+1}}l_{\tau+1}\frac{\P
                                            f_{l_{\tau+1},a_\tau,h_\tau}}{\P
                                            f_{a_s, h_s}}\prod_{k=1}^\tau\frac{\P f_{\d_k, h_k}}{\P f_{a_k,
                                           h_k}}\\
                                         &=-\sum_{
                                           \underline a_{s+1},
                                           \underline
                                           h_{s+1}}\Q_\tau(a_\tau, h_\tau)\frac{\P
                                           f_{a_\tau,h_\tau}}{\P
                                           f_{a_s, h_s}}\prod_{k=1}^\tau\frac{\P f_{\d_k, h_k}}{\P f_{a_k,
                                           h_k}}\\
                                         &=-\left(\sum_{
                                           \underline a_{s+1},
                                           \underline
                                           h_{s+1}}\Q_\tau(a_\tau, h_\tau)\frac{\P
                                           f_{a_\tau,h_\tau}}{\P
                                           f_{a_s,
                                           h_s}}\right)\left(\prod_{k=1}^\tau\r(a_k,
                                           h_k)\right)\\
                                            &= \Q_s(a_s, h_s)\prod_{k=1}^s\r(a_k,
                                              h_k), 
\end{align*}
where we used the fact that for continuous variables under Assumption~\ref{ass:inv} we can
write
\[\Q_s(a_s, h_s)=\sum_{\underline a_{s+1},
    \underline h_{s+1}}\Q_\tau(a_\tau,
  h_\tau)\prod_{k=s+1}^\tau\g(a_k\mid h_k) \prod_{k=s+1}^{\tau}\r(a_k,
  h_k),\] using the change of variable formula (the above display
holds trivially for discrete $A_t$). Notice now that
\[\sum_{f\in\mathcal{F}}\frac{\dd\hat \Theta^*(\P)}{\dd\P f}\P
  f=\Theta(P)\]
for all functions in $\mathcal F$, and that
\begin{align*}
  \sum_{l_{\tau+1},a_\tau,h_\tau}\frac{\dd \Theta(\P)}{\dd f_{l_{\tau+1},a_\tau,h_\tau}}
  f_{l_{\tau+1},a_\tau,h_\tau}(Z) &=
                                    l_{\tau+1}\prod_{k=1}^\tau\r(A_k, H_k)\\
  \sum_{a_s,h_s} \frac{\dd \Theta(\P)}{\dd \P f_{\d_s,h_s}}f_{\d_s,h_s}(Z) &= \Q_s(A^\d_s, H_s)\prod_{k=1}^{s-1}\r(A_k,
  H_k)\\
  \sum_{a_s,h_s} \frac{\dd \Theta(\P)}{\dd \P f_{a_s,h_s}}f_{a_s,h_s}(Z) &=\Q_s(A_s, H_s)\prod_{k=1}^s\r(A_k,
                                              H_k).
\end{align*}
Putting these results together and summing over $s$ yields the
conjectured $\dr_P(Z)=\D_1(Z;\eta)-\Theta(P)$. Then, we can confirm that
$\D_1(Z;\eta)$ satisfies (\ref{eq:defeif}). To see this, note that
Lemma~\ref{theo:sdr} implies that the functional satisfies the
following expansion:
\[ \Theta(\P_\epsilon) = \Theta(\P) + \int \{\D_1(Z;\eta) - \Theta(\P)\} \dd\P_\epsilon - \rem_0(\eta_\epsilon,\eta),\]
where we denote
\[ \rem_0(\eta',\eta) = \sum_{s=1}^\tau\E\left[\{\r_s'(A_s, H_s) -
    \r_s(A_s, H_s)\}\{\Q_s'(A_s, H_s) - \Q_s(A_s, H_s)\} \right].\]
Differentiating with respect to $\epsilon$ and evaluating at
$\epsilon=0$ yields
\begin{align*}
  \frac{\dd }{\dd \epsilon}\Theta(\P_\epsilon)\bigg|_{\epsilon = 0}
  &= \int\{\D_1(Z;\eta) - \Theta(\P)\}
    \left(\frac{\dd  \P_\epsilon}{\dd \epsilon}\right)\bigg|_{\epsilon
    = 0} - \frac{\dd }{\dd
    \epsilon}\rem_0(\eta_\epsilon,\eta)\bigg|_{\epsilon = 0} \\
&= \int\{\D_1(Z;\eta) - \Theta(\P)\}
    \left(\frac{\dd \log \P_\epsilon}{\dd
    \epsilon}\right)\bigg|_{\epsilon = 0}\dd P - \frac{\dd }{\dd \epsilon}\rem_0(\eta_\epsilon,\eta)\bigg|_{\epsilon = 0},
\end{align*}
and the result follows after noticing that
\[\frac{\dd }{\dd \epsilon}\rem_0(\eta_\epsilon,\eta)\bigg|_{\epsilon = 0}=0.\]
\end{proof}

\section{Sequential double robustness (Lemma \ref{theo:sdr})}
\begin{proof}
  This lemma follows from recursive application for
  $s=t+1,\ldots,\tau$ of the following
  relationship:
  \begin{align*}
    \textcolor{blue}{\E[\Q_{s,1}}&\textcolor{blue}{(A^\d_s, H_s) - \Q_{s}(A^\d_s, H_s)\mid A_{s-1},
    H_{s-1}]} =\\
    = -&\E\left\{\frac{\g_{s,1}^\d(A_s\mid
               H_s)}{\g_{s,1}(A_s\mid H_s)}[\Q_{s+1, 1}(A^\d_{s+1},
         H_{s+1}) - \Q_{s, 1}(A_s,
         H_s)]\,\,\bigg|\,\,A_{s-1},H_{s-1}\right\}\\
    +&\E\left\{\left[\frac{\g_s^\d(A_s\mid
       H_s)}{\g_s(A_s\mid H_s)}-\frac{\g_{s,1}^\d(A_s\mid
       H_s)}{\g_{s,1}(A_s\mid H_s)}\right][\Q_{s,1}(A_s,
       H_s) - \Q_s(A_s,
       H_s)]\,\,\bigg|\,\,A_{s-1},H_{s-1}\right\}\\
    +&\E\left\{\frac{\g_{s,1}^\d(A_s\mid
       H_s)}{\g_{s,1}(A_s\mid H_s)}\textcolor{blue}{\E[\Q_{s+1, 1}(A^\d_{s+1},
       H_{s+1}) - \Q_{s+1}(A^\d_{s+1},
       H_{s+1})\mid A_s, H_s]}\,\,\bigg|\,\,A_{s-1},H_{s-1}\right\},
  \end{align*}
  which follows from application of the change of variable formula
  under Assumption~\ref{ass:inv}.
\end{proof}

\section{Second order representation for SDR
  (Lemma~\ref{lemma:remorder})}
In this proof we denote $\check\eta =
(\hat\r_1,\check\Q_1,\ldots,\hat\r_\tau,\check\Q_\tau)$, as well as
\[\rem_t^\dagger(a_t,
  h_t;\check\eta)=\sum_{s=t+1}^{\tau}\E\left[\hat C_{t,s}\{\hat\r_s(A_s,
    H_s)-\r_s(A_s, H_s)\}\{\check\Q_s(A_s,
    H_s)-\check\Q^\dagger_s(A_s, H_s)\}\mid A_t=a_t, H_t=h_t\right],\]
where
\[\hat C_{t, s} = \prod_{r=t+1}^{s-1}\hat \r_r(A_r,H_r).\]
We simplify the notation by omitting the arguments of the
functions whenever they are clear from context. First of all, note
that plugging in equation (\ref{eq:check}) into equation (\ref{eq:fo})
yields
\[\rem_t(\check\eta) = \rem_t^\dagger(\check\eta) +
  \sum_{s=t+1}^{\tau}\E\left[(\r_s-\hat\r_s)\rem_s(\check\eta)\mid
    a_t, h_t\right].\] Applying the above equality inductively for
$t=\tau-1,\ldots,0$ yields
\[\rem_0(\check\eta) = \rem_0^\dagger(\check\eta)
  +\sum_{s=1}^{\tau}\sum_{m=1}^s\E\left[\prod_{k=m}^s\hat C_{0,s}(
    \r_k-\hat\r_k)\rem_s^\dagger(\check\eta)\right],\]
where we note that under the assumption of the lemma
\[\E\left[\prod_{k=m}^s(
    \r_k-\hat\r_k)\rem_s^\dagger(\check\eta)\right]\leq
  C_1 \E[\rem_s^\dagger(\check\eta)]|,\]
for a constant $C_1$. The result follows from noticing that
\[\E[\rem_s^\dagger(\check\eta)]\leq C_2\sum_{k=s+1}^\tau||\hat\r_k -
  \r_k||\,||\check\Q_k - \check\Q_k^\dagger||,\]which follows from the law of iterated expectation and the
Cauchy-Schwarz inequality.
\section{Asymptotic Normality of TMLE (Theorem~\ref{theo:astmle})}
\begin{proof}
  Let $\Pnj$ denote the empirical distribution of the prediction set
  ${\cal V}_j$, and let $\Gnj$ denote the associated empirical process
  $\sqrt{n/J}(\Pnj-\P)$. Let $\Gn$ denote the empirical process
  $\sqrt{n}(\Pn-\P)$.  We use $E(g(Z_1,\ldots,Z_n))$ to denote
  expectation with respect to the joint distribution of
  $(Z_1,\ldots,Z_n)$ (as opposed to the script letter $\E$ used to
  denote $\E(f(Z))=\int f(z)\dd\P(z)$ in the main manuscript). In
  this proof we use the alternative notation
  $\D_\eta(Z)=\D_1(Z;\eta)$. By definition of the TMLE and the fact
  that it solves the efficient influence function estimating equation,
  we have
\[\thetatmle = \frac{1}{J}\sum_{j=1}^J\Pnj \D_{\tilde \eta_j}.\]
Thus,
\begin{equation}
  \sqrt{n}(\thetatmle - \theta)=\Gn (\D_\eta
  - \theta) + R_{n,1} + R_{n,2},\label{eq:exptmle}
\end{equation}
where
\[  R_{n,1}  =\frac{1}{\sqrt{J}}\sum_{j=1}^J\Gnj(\D_{\tilde
    \eta_j} - \D_{\eta}),\,\,\,
  R_{n,2}  = \frac{\sqrt{n}}{J}\sum_{j=1}^J\P(\D_{\tilde
    \eta_j}-\theta).
\]
Lemma~\ref{theo:sdr} together with the Cauchy-Schwartz inequality and
the assumptions of the theorem shows that $R_{n,2}=o_P(1)$.

Note that $\D_{\tilde \eta_j}$ depends on the full sample through the
estimates of the parameters $\epsilon$ of the generalized linear
tilting models. To make this dependence explicit, we introduce the
notation $\D_{\hat \eta_j,\hat\epsilon}=\D_{\tilde \eta_j}$ and
$R_{n,1}(\epsilon)$. Note that, since $\hat\eta$ is consistent, we can
find a deterministic sequence $\delta_n\to 0$ satisfying
$P(\hat\epsilon<\delta_n)\to 1$ ($\hat\epsilon$ converges to 0 in probability). Let
${\cal F}_n^j=\{\D_{\hat \eta_j,\epsilon} -
\D_\eta:\epsilon<\delta_n\}$. Because the function $\hat\eta_j$ is
fixed given the training data, we can apply Theorem 2.14.2 of
\cite{vanderVaart&Wellner96} to obtain
\begin{equation}
  E\left\{\sup_{f\in {\cal F}_n^j}|\Gnj f| \,\,\bigg|\,\, {\cal
      T}_j\right\}\lesssim \lVert F^j_n \rVert \int_0^1
  \sqrt{1+N_{[\,]}(\alpha \lVert F_n^j \rVert, {\cal F}_n^j,
    L_2(\P))}\dd\alpha,\label{eq:empproc}
\end{equation} where
$N_{[\,]}(\alpha \lVert F_n^j \rVert, {\cal F}_n^j, L_2(\P))$ is the
bracketing number and we take
$F_n^j=\sup_{\epsilon<\delta_n} \lvert \D_{\hat \eta_j,\epsilon} - \D_\eta \rvert$
as an envelope for the class ${\cal F}_n^j$. Theorem 2.7.2 of \cite{vanderVaart&Wellner96} shows
\[\log N_{[\,]}(\alpha \lVert F_n^j \rVert, {\cal F}_n^j, L_2(\P))\lesssim
  \frac{1}{\alpha \lVert F_n^j \rVert}.\]
This shows
\begin{align*}
  \lVert F^j_n \rVert \int_0^1\sqrt{1+N_{[\,]}(\alpha
  \lVert F_n^j \rVert, {\cal F}_n^j, L_2(\P))}\dd\alpha &\lesssim
                                                            \int_0^1\sqrt{\lVert F^j_n \rVert^2+\frac{\lVert F^j_n
                                                            \rVert}{\alpha}}\dd\alpha\\ &\leq
                                                                                              \lVert F^j_n \rVert + \lVert F^j_n \rVert^{1/2}
                                                                                              \int_0^1\frac{1}{\alpha^{1/2}}\dd\alpha\\
                                                          &\leq \lVert F^j_n \rVert + 2 \lVert F^j_n \rVert^{1/2}.
\end{align*}
Since $\hat\eta$ is consistent and $\delta_n\to 0$, $\lVert F^j_n \rVert = o_P(1)$. This
shows $\sup_{f\in {\cal F}_n^j}\Gnj f=o_P(1)$ for each $j$,
conditional on ${\cal T}_j$. Thus $
R_{n,1}=o_P(1)$. Lemma~\ref{theo:sdr} together with the
Cauchy-Schwartz inequality and the assumptions of the theorem shows
that $R_{n,2}=o_P(1)$, concluding the proof of the theorem.
\end{proof}

\section{Other results}
\begin{lemma}\label{lemma:altrep}
  Under Assumption~\ref{ass:inv}, we have
  \[ E[Y(\bar A^\d)] = \int_{\mathcal {\bar A}_\tau, \mathcal {\bar
        L}_\tau}\E[Y\mid A_\tau = a_\tau, H_\tau =
    h_\tau]\prod_{k=1}^\tau\g_k^\d(a_k\mid h_k) \dd
    \nu(a_k)\dd\P(l_k\mid a_{k-1},h_{k-1}),\] where $\g^\d$ is defined
  in (\ref{eq:gdelta}) in the main document.
\end{lemma}

\begin{proof}
  This follows from recursive application for $t=\tau+1,\ldots,1$ of the
  change of variable formula to (\ref{eq:eq1}) under Assumption~\ref{ass:inv}.
\end{proof}

\section{Supplemental Figures}

\begin{figure}[H]
  \centering
  \includegraphics[scale = 0.8]{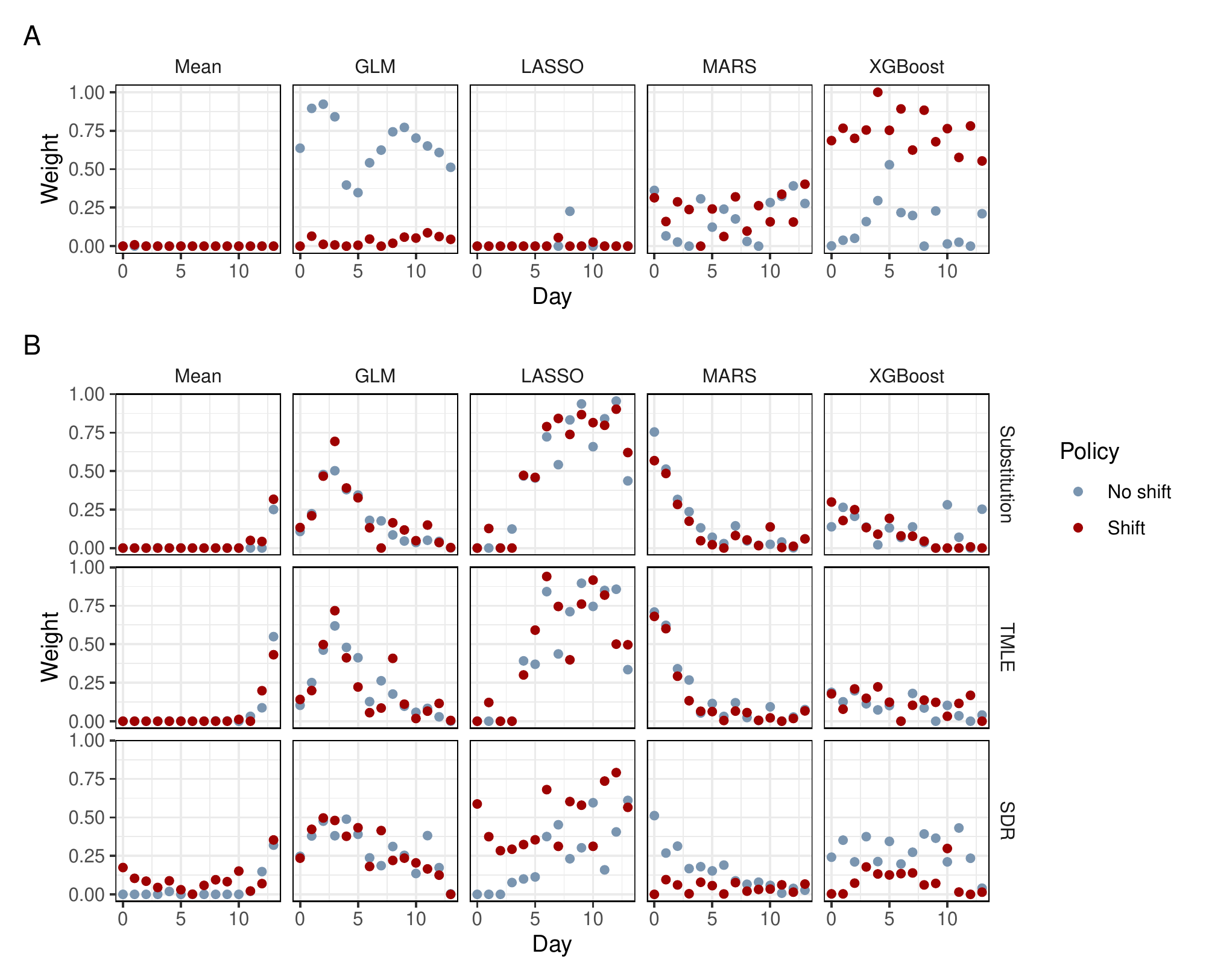}
  \caption{Weights (average across cross-fitting folds) of each
    learner for estimation of $\r_t$ (panel A) and $\Q_t$ (panel B) at
    each time point in the illustrative study on P/F ratios and 28-day
    mortality among ICU patients.}\label{fig:weights}
\end{figure}

\begin{figure}[H]
  \centering
  \includegraphics[scale = 0.5]{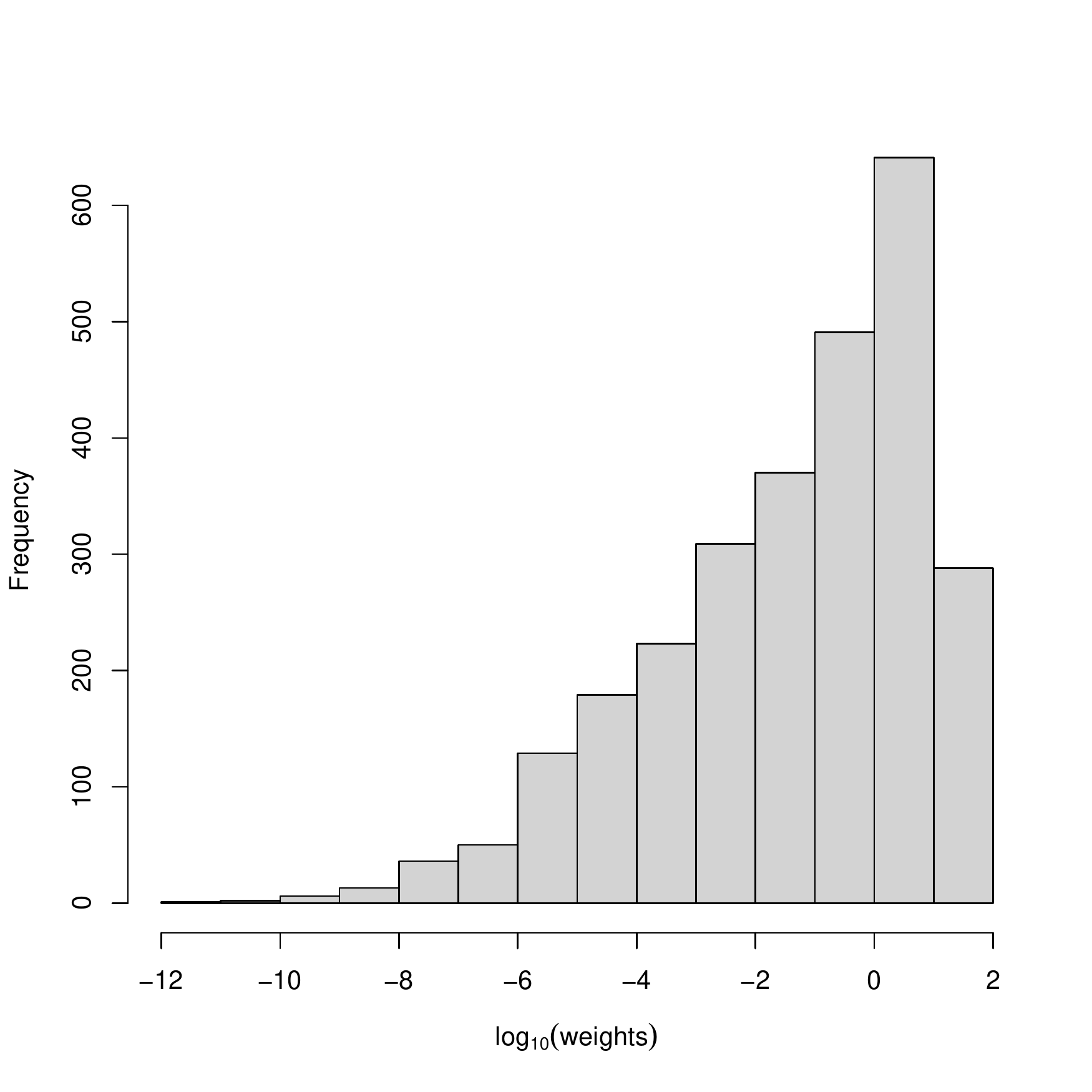}
  \caption{Histogram of
    $\log_{10}(\prod_{t=1}^{14}\hat\r_t(A_t, H_t))$ in the
    illustrative example. Values where
    $\prod_{t=1}^{14}\hat\r_t(A_t, H_t) = 0$ are
    omitted.}\label{fig:ipw}
\end{figure}

\bibliographystyle{plainnat}
\bibliography{refs}

%% file: _preamble.tex
\usepackage[inline,shortlabels]{enumitem}
\usepackage{float}
\usepackage{mathtools}
\usepackage{graphicx}
\usepackage[round]{natbib}
\usepackage[text={6in,9.3in}]{geometry}
\usepackage{tikz}
\usepackage[english]{babel}
\usepackage{longtable}
\usepackage{color}
\usepackage{soul,xcolor}
\usepackage[pdftex]{changebar}
\usepackage{bbm}
\usepackage{amssymb,amsmath,amsthm}
\usepackage{multirow}
\usepackage[titletoc,title]{appendix}
\usepackage{authblk}
\usepackage{setspace}
\usepackage{dsfont}
\usepackage{booktabs}
\usepackage{multirow}
\usepackage{caption}
\usepackage[OT1]{fontenc}
\usepackage{subcaption}
\usepackage{refcount}
\usepackage{accents}
\usepackage{setspace}
\usepackage{thmtools}
\usepackage{changepage} 

\graphicspath{ {../simulations/01_init_manuscript/graphs/} }

\usepackage{xr-hyper}
\usepackage[colorlinks,citecolor=blue,urlcolor=blue]{hyperref}

\usepackage[inline]{trackchanges}
\addeditor{Ivan}
\addeditor{Nima}

\newtheorem{proposition}{Proposition}
\newtheorem{theorem}{Theorem}

\theoremstyle{definition}

\AtEndDocument{\refstepcounter{theorem}\label{finalthm}}
\AtEndDocument{\refstepcounter{equation}\label{finaleq}}

{
  \theoremstyle{definition}
  \newtheorem{assumption}{Assumption}{}
}
{
  \theoremstyle{definition}
  
}
{
  \theoremstyle{definition}
  \newtheorem{example}{Example}[section]
}

\newcommand{\supp}{\mathop{\mathrm{supp}}}

\newtheorem{lemma}{Lemma} 
\newtheorem{definition}{Definition}

\DeclareMathOperator{\logit}{logit} \DeclareMathOperator{\var}{\mathsf{Var}}
\DeclareMathOperator{\link}{link}
 
\renewcommand{\P}{\mathsf{P}}
\newcommand{\Q}{\mathsf{m}}
\newcommand{\rem}{\mathsf{Rem}}
\newcommand{\D}{\mathsf{\phi}}

\renewcommand{\d}{\mathbbm{d}}
\renewcommand{\b}{\mathbbm{b}}
\newcommand{\p}{\mathsf{p}}

\renewcommand{\r}{\mathsf{r}}
\newcommand{\g}{\mathsf{g}}
\newcommand{\G}{\mathsf{G}}

\newcommand{\uu}{\mathsf{u}}

\newcommand{\indep}{\mbox{$\perp\!\!\!\perp$}} 
 \newcommand{\dd}{\mathrm{d}}

\newcommand{\Pn}{\mathsf{P}_n}

\newcommand{\thetalmtp}{\theta_{\mbox{\scriptsize lmtp}}}
\newcommand{\thetalsi}{\theta_{\mbox{\scriptsize lmtp-si}}}
\newcommand{\thetasub}{\hat\theta_{\mbox{\scriptsize sub}}}
\newcommand{\thetaipw}{\hat\theta_{\mbox{\scriptsize ipw}}}
\newcommand{\thetatmle}{\hat\theta_{\mbox{\scriptsize tmle}}}
\newcommand{\thetasr}{\hat\theta_{\mbox{\scriptsize sr}}}
\newcommand{\one}{\mathds{1}}

\newcommand{\E}{\mathsf{E}}

\DeclarePairedDelimiterX{\norm}[1]{\lVert}{\rVert}{#1}

\usepackage{ulem}
\newcommand\redsout{\bgroup\markoverwith{\textcolor{red}{\rule[0.5ex]{2pt}{2.0pt}}}\ULon}

\pgfdeclarelayer{background}
\pgfsetlayers{background,main}
\usetikzlibrary{arrows,positioning}
\tikzset{
  >=stealth',
  punkt/.style={
    rectangle,
    rounded corners,
    draw=black, very thick,
    text width=6.5em,
    minimum height=2em,
    text centered},
  pil/.style={
    ->,
    thick,
    shorten <=2pt,
    shorten >=2pt,}
}
\newcommand{\Vertex}[2]
{\node[minimum width=0.6cm,inner sep=0.05cm] (#2) at (#1) {$\footnotesize#2$};
}
\newcommand{\Vertexr}[2]
{\node[rectangle, draw, minimum width=0.6cm,inner sep=0.05cm] (#2) at (#1) {$\footnotesize#2$};
}
\newcommand{\ArrowR}[3]%
{ \begin{pgfonlayer}{background}
    \draw[->,#3] (#1) to[bend right=30] (#2);
  \end{pgfonlayer}
}
\newcommand{\ArrowL}[3]%
{ \begin{pgfonlayer}{background}
    \draw[->,#3] (#1) to[bend left=45] (#2);
  \end{pgfonlayer}
}
\newcommand{\EdgeL}[3]%
{ \begin{pgfonlayer}{background}
    \draw[dashed,#3] (#1) to[bend right=-45] (#2);
  \end{pgfonlayer}
}

\newcommand{\Arrow}[3]%
{ \begin{pgfonlayer}{background}
    \draw[->,#3] (#1) -- +(#2);
  \end{pgfonlayer}
}
\newcommand{\titlepaper}{Non-parametric causal effects based on
  longitudinal modified treatment policies}

\date{\today}
\pdfoutput=1

\ifx\isblinded\undefined
\author[1]{Iv\'an D\'iaz \thanks{corresponding author:
    ild2005@med.cornell.edu}}

\author[1]{Nicholas Williams}
\author[1]{Katherine L. Hoffman}
\author[2]{Edward J. Schenck}

\affil[1]{\small Division of Biostatistics, Department of Population
  Health Sciences, Weill Cornell Medicine.}
\affil[2]{\small Division of Pulmonary \& Critical Care Medicine, Department of Medicine, Weill Cornell Medicine.}
\else
\if\isblinded1
\author{}
\else
\author[1]{Iv\'an D\'iaz \thanks{corresponding author:
    ild2005@med.cornell.edu}}

\author[1]{Nicholas Williams}
\author[1]{Katherine L. Hoffman}
\author[2]{Edward J. Schenck}

\affil[1]{\small Division of Biostatistics, Department of Population
  Health Sciences, Weill Cornell Medicine.}
\affil[2]{\small Division of Pulmonary \& Critical Care Medicine, Department of Medicine, Weill Cornell Medicine.}
\fi
\fi

\ifx\arxiv\undefined
\else
\fi

%% file: paper_arxiv.bbl
\begin{thebibliography}{55}
\providecommand{\natexlab}[1]{#1}
\providecommand{\url}[1]{\texttt{#1}}
\expandafter\ifx\csname urlstyle\endcsname\relax
  \providecommand{\doi}[1]{doi: #1}\else
  \providecommand{\doi}{doi: \begingroup \urlstyle{rm}\Url}\fi

\bibitem[Ai and Chen(2003)]{ai2003efficient}
Chunrong Ai and Xiaohong Chen.
\newblock Efficient estimation of models with conditional moment restrictions
  containing unknown functions.
\newblock \emph{Econometrica}, 71\penalty0 (6):\penalty0 1795--1843, 2003.

\bibitem[Bang and Robins(2005)]{Bang05}
Heejung Bang and James~M Robins.
\newblock Doubly robust estimation in missing data and causal inference models.
\newblock \emph{Biometrics}, 61\penalty0 (4):\penalty0 962--973, 2005.

\bibitem[Benkeser and van~der Laan(2016)]{benkeser2016highly}
David Benkeser and Mark van~der Laan.
\newblock The highly adaptive lasso estimator.
\newblock In \emph{2016 IEEE International Conference on Data Science and
  Advanced Analytics (DSAA)}, pages 689--696. IEEE, 2016.

\bibitem[Benkeser et~al.(2016)Benkeser, Carone, van~der Laan, and
  Gilbert]{benkeser2016doubly}
David Benkeser, Marco Carone, Mark~J van~der Laan, and Peter Gilbert.
\newblock Doubly-robust nonparametric inference on the average treatment
  effect.
\newblock Technical Report 356, U.C. Berkeley Division of Biostatistics Working
  Paper Series, 2016.

\bibitem[Bickel et~al.(1997)Bickel, Klaassen, Ritov, and Wellner]{Bickel97}
Peter~J Bickel, Chris~AJ Klaassen, YA'Acov Ritov, and Jon~A Wellner.
\newblock \emph{Efficient and Adaptive Estimation for Semiparametric Models}.
\newblock Springer-Verlag, 1997.

\bibitem[Bickel et~al.(2009)Bickel, Ritov, Tsybakov,
  et~al.]{bickel2009simultaneous}
Peter~J Bickel, Ya’acov Ritov, Alexandre~B Tsybakov, et~al.
\newblock Simultaneous analysis of lasso and dantzig selector.
\newblock \emph{The Annals of Statistics}, 37\penalty0 (4):\penalty0
  1705--1732, 2009.

\bibitem[Breiman(1996)]{Breiman1996}
Leo Breiman.
\newblock Stacked regressions.
\newblock \emph{Machine learning}, 24\penalty0 (1):\penalty0 49--64, 1996.

\bibitem[Buckley and James(1979)]{buckley1979linear}
Jonathan Buckley and Ian James.
\newblock Linear regression with censored data.
\newblock \emph{Biometrika}, 66\penalty0 (3):\penalty0 429--436, 1979.

\bibitem[Chen and White(1999)]{chen1999improved}
Xiaohong Chen and Halbert White.
\newblock Improved rates and asymptotic normality for nonparametric neural
  network estimators.
\newblock \emph{IEEE Transactions on Information Theory}, 45\penalty0
  (2):\penalty0 682--691, 1999.

\bibitem[Cheng et~al.(2004)Cheng, Chu, et~al.]{cheng2004semiparametric}
Kuang~Fu Cheng, Chih-Kang Chu, et~al.
\newblock Semiparametric density estimation under a two-sample density ratio
  model.
\newblock \emph{Bernoulli}, 10\penalty0 (4):\penalty0 583--604, 2004.

\bibitem[Chernozhukov et~al.(2018)Chernozhukov, Chetverikov, Demirer, Duflo,
  Hansen, Newey, and Robins]{chernozhukov2018double}
Victor Chernozhukov, Denis Chetverikov, Mert Demirer, Esther Duflo, Christian
  Hansen, Whitney Newey, and James Robins.
\newblock Double/debiased machine learning for treatment and structural
  parameters.
\newblock \emph{The Econometrics Journal}, 21\penalty0 (1):\penalty0 C1--C68,
  2018.

\bibitem[D{\'\i}az(2019)]{diaz2019statistical}
Iv{\'a}n D{\'\i}az.
\newblock Statistical inference for data-adaptive doubly robust estimators with
  survival outcomes.
\newblock \emph{Statistics in Medicine}, 38\penalty0 (15):\penalty0 2735--2748,
  2019.

\bibitem[D{\'i}az and Hejazi(2020)]{diaz2020}
Iv{\'a}n D{\'i}az and Nima~S. Hejazi.
\newblock Causal mediation analysis for stochastic interventions.
\newblock \emph{Journal of the Royal Statistical Society: Series B (Statistical
  Methodology)}, n/a\penalty0 (n/a), 2020.
\newblock \doi{10.1111/rssb.12362}.
\newblock URL
  \url{https://rss.onlinelibrary.wiley.com/doi/abs/10.1111/rssb.12362}.

\bibitem[D\'iaz and {van der Laan}(2012)]{Diaz12}
Iv\'an D\'iaz and Mark~J {van der Laan}.
\newblock Population intervention causal effects based on stochastic
  interventions.
\newblock \emph{Biometrics}, 68\penalty0 (2):\penalty0 541--549, 2012.

\bibitem[D{\'\i}az and van~der Laan(2013{\natexlab{a}})]{diaz2013assessing}
Iv{\'a}n D{\'\i}az and Mark~J van~der Laan.
\newblock Assessing the causal effect of policies: an example using stochastic
  interventions.
\newblock \emph{The international journal of biostatistics}, 9\penalty0
  (2):\penalty0 161--174, 2013{\natexlab{a}}.

\bibitem[D{\'\i}az and van~der Laan(2013{\natexlab{b}})]{diaz2013targeted}
Iv{\'a}n D{\'\i}az and Mark~J van~der Laan.
\newblock Targeted data adaptive estimation of the causal dose--response curve.
\newblock \emph{Journal of Causal Inference}, 1\penalty0 (2):\penalty0
  171--192, 2013{\natexlab{b}}.

\bibitem[D{\'\i}az and van~der Laan(2017)]{diaz2017doubly}
Iv{\'a}n D{\'\i}az and Mark~J van~der Laan.
\newblock Doubly robust inference for targeted minimum loss--based estimation
  in randomized trials with missing outcome data.
\newblock \emph{Statistics in medicine}, 36\penalty0 (24):\penalty0 3807--3819,
  2017.

\bibitem[D{\'\i}az and {van der Laan}(2018)]{diaz2018stochastic}
Iv{\'a}n D{\'\i}az and Mark~J {van der Laan}.
\newblock Stochastic treatment regimes.
\newblock In \emph{Targeted Learning in Data Science}, pages 219--232.
  Springer, 2018.

\bibitem[Farrell(2015)]{farrell2015robust}
Max~H Farrell.
\newblock Robust inference on average treatment effects with possibly more
  covariates than observations.
\newblock \emph{Journal of Econometrics}, 189\penalty0 (1):\penalty0 1--23,
  2015.

\bibitem[Foster and Syrgkanis(2019)]{foster2019orthogonal}
Dylan~J Foster and Vasilis Syrgkanis.
\newblock Orthogonal statistical learning.
\newblock \emph{arXiv preprint arXiv:1901.09036}, 2019.

\bibitem[Gruber and van~der Laan(2010)]{Gruber2010t}
Susan Gruber and Mark~J van~der Laan.
\newblock A targeted maximum likelihood estimator of a causal effect on a
  bounded continuous outcome.
\newblock \emph{The International Journal of Biostatistics}, 6\penalty0 (1),
  2010.

\bibitem[Haneuse and Rotnitzky(2013)]{Haneuse2013}
Sebastian Haneuse and Andrea Rotnitzky.
\newblock Estimation of the effect of interventions that modify the received
  treatment.
\newblock \emph{Statistics in Medicine}, 2013.

\bibitem[Kennedy(2019)]{kennedy2018nonparametric}
Edward~H Kennedy.
\newblock Nonparametric causal effects based on incremental propensity score
  interventions.
\newblock \emph{Journal of the American Statistical Association}, 114\penalty0
  (526):\penalty0 645--656, 2019.

\bibitem[Kennedy(2020)]{kennedy2020optimal}
Edward~H Kennedy.
\newblock Optimal doubly robust estimation of heterogeneous causal effects.
\newblock \emph{arXiv preprint arXiv:2004.14497}, 2020.

\bibitem[Kennedy et~al.(2017)Kennedy, Ma, McHugh, and Small]{kennedy2017non}
Edward~H Kennedy, Zongming Ma, Matthew~D McHugh, and Dylan~S Small.
\newblock Non-parametric methods for doubly robust estimation of continuous
  treatment effects.
\newblock \emph{Journal of the Royal Statistical Society: Series B (Statistical
  Methodology)}, 79\penalty0 (4):\penalty0 1229--1245, 2017.

\bibitem[Klaassen(1987)]{klaassen1987consistent}
Chris~AJ Klaassen.
\newblock Consistent estimation of the influence function of locally
  asymptotically linear estimators.
\newblock \emph{The Annals of Statistics}, pages 1548--1562, 1987.

\bibitem[Luedtke et~al.(2017)Luedtke, Sofrygin, van~der Laan, and
  Carone]{luedtke2017sequential}
Alexander~R Luedtke, Oleg Sofrygin, Mark~J van~der Laan, and Marco Carone.
\newblock Sequential double robustness in right-censored longitudinal models.
\newblock \emph{arXiv preprint arXiv:1705.02459}, 2017.

\bibitem[Molina et~al.(2017)Molina, Rotnitzky, Sued, and
  Robins]{molina2017multiple}
J~Molina, A~Rotnitzky, M~Sued, and JM~Robins.
\newblock Multiple robustness in factorized likelihood models.
\newblock \emph{Biometrika}, 104\penalty0 (3):\penalty0 561--581, 2017.

\bibitem[Neugebauer and van~der Laan(2007)]{Neugebauer2007419}
R.~Neugebauer and M.~J. van~der Laan.
\newblock Nonparametric causal effects based on marginal structural models.
\newblock \emph{Journal of Statistical Planning \& Inference}, 137\penalty0
  (2):\penalty0 419 -- 434, 2007.
\newblock ISSN 0378-3758.
\newblock \doi{DOI: 10.1016/j.jspi.2005.12.008}.

\bibitem[Pearl(2009)]{Pearl00}
Judea Pearl.
\newblock \emph{Causality: Models, Reasoning, and Inference}.
\newblock Cambridge University Press, Cambridge, 2009.

\bibitem[Qin(1998)]{qin1998inferences}
Jing Qin.
\newblock Inferences for case-control and semiparametric two-sample density
  ratio models.
\newblock \emph{Biometrika}, 85\penalty0 (3):\penalty0 619--630, 1998.

\bibitem[Richardson and Robins(2013)]{richardson2013single}
Thomas~S Richardson and James~M Robins.
\newblock Single world intervention graphs (swigs): A unification of the
  counterfactual and graphical approaches to causality.
\newblock \emph{Center for the Statistics and the Social Sciences, University
  of Washington Series. Working Paper}, 128\penalty0 (30):\penalty0 2013, 2013.

\bibitem[Robins et~al.(2009)Robins, Li, Tchetgen, and van~der
  Vaart]{robins2009quadratic}
James Robins, Lingling Li, Eric Tchetgen, and Aad~W van~der Vaart.
\newblock Quadratic semiparametric von mises calculus.
\newblock \emph{Metrika}, 69\penalty0 (2-3):\penalty0 227--247, 2009.

\bibitem[Robins et~al.(2004)Robins, Hern{\'a}n, and SiEBERT]{robins2004effects}
James~M Robins, Miguel~A Hern{\'a}n, and UWE SiEBERT.
\newblock Effects of multiple interventions.
\newblock \emph{Comparative quantification of health risks: global and regional
  burden of disease attributable to selected major risk factors}, 1:\penalty0
  2191--2230, 2004.

\bibitem[Robins(2000)]{Robins00}
J.M. Robins.
\newblock Robust estimation in sequentially ignorable missing data and causal
  inference models.
\newblock In \emph{Proceedings of the American Statistical Association}, 2000.

\bibitem[Robins et~al.(1994)Robins, Rotnitzky, and
  Zhao]{Robins&Rotnitzky&Zhao94}
J.M. Robins, A.~Rotnitzky, and L.P. Zhao.
\newblock Estimation of regression coefficients when some regressors are not
  always observed.
\newblock \emph{Journal of the American Statistical Association}, 89\penalty0
  (427):\penalty0 846--866, September 1994.

\bibitem[Rotnitzky et~al.(2017)Rotnitzky, Robins, and
  Babino]{rotnitzky2017multiply}
Andrea Rotnitzky, James Robins, and Lucia Babino.
\newblock On the multiply robust estimation of the mean of the g-functional.
\newblock \emph{arXiv preprint arXiv:1705.08582}, 2017.

\bibitem[Rubin and van~der Laan(2005)]{rubin2005general}
Dan Rubin and Mark~J van~der Laan.
\newblock A general imputation methodology for nonparametric regression with
  censored data.
\newblock 2005.

\bibitem[Rubin and van~der Laan(2007)]{rubin2007doubly}
Daniel Rubin and Mark~J van~der Laan.
\newblock A doubly robust censoring unbiased transformation.
\newblock \emph{The international journal of biostatistics}, 3\penalty0 (1),
  2007.

\bibitem[Smucler et~al.(2019)Smucler, Rotnitzky, and
  Robins]{smucler2019unifying}
Ezequiel Smucler, Andrea Rotnitzky, and James~M Robins.
\newblock A unifying approach for doubly-robust $\ell_1$ regularized estimation
  of causal contrasts.
\newblock \emph{arXiv preprint arXiv:1904.03737}, 2019.

\bibitem[Stock(1989)]{stock1989nonparametric}
James~H Stock.
\newblock Nonparametric policy analysis.
\newblock \emph{Journal of the American Statistical Association}, 84\penalty0
  (406):\penalty0 567--575, 1989.

\bibitem[Taubman et~al.(2009)Taubman, Robins, Mittleman, and
  Hern{\'a}n]{taubman2009intervening}
Sarah~L Taubman, James~M Robins, Murray~A Mittleman, and Miguel~A Hern{\'a}n.
\newblock Intervening on risk factors for coronary heart disease: an
  application of the parametric g-formula.
\newblock \emph{International journal of epidemiology}, 38\penalty0
  (6):\penalty0 1599--1611, 2009.

\bibitem[{Tchetgen Tchetgen}(2009)]{tchetgen2009commentary}
Eric~J {Tchetgen Tchetgen}.
\newblock A commentary on g. molenberghs's review of missing data methods.
\newblock \emph{Drug Information Journal}, 43\penalty0 (4):\penalty0 433--435,
  2009.

\bibitem[van~der Laan(2014)]{van2014targeted}
Mark~J van~der Laan.
\newblock Targeted estimation of nuisance parameters to obtain valid
  statistical inference.
\newblock \emph{The international journal of biostatistics}, 10\penalty0
  (1):\penalty0 29--57, 2014.

\bibitem[van~der Laan and Gruber(2012)]{van2012targeted}
Mark~J van~der Laan and Susan Gruber.
\newblock Targeted minimum loss based estimation of causal effects of multiple
  time point interventions.
\newblock \emph{The international journal of biostatistics}, 8\penalty0 (1),
  2012.

\bibitem[{van der Laan} and Robins(2003)]{vanderLaan2003}
Mark~J {van der Laan} and James~M Robins.
\newblock \emph{Unified Methods for Censored Longitudinal Data and Causality}.
\newblock Springer, New York, 2003.

\bibitem[{van der Laan} and Rose(2011)]{vanderLaanRose11}
Mark~J {van der Laan} and Sherri Rose.
\newblock \emph{Targeted Learning: Causal Inference for Observational and
  Experimental Data}.
\newblock Springer, New York, 2011.

\bibitem[{van der Laan} and Rose(2018)]{vanderLaanRose18}
Mark~J {van der Laan} and Sherri Rose.
\newblock \emph{Targeted Learning in Data Science: Causal Inference for Complex
  longitudinal Studies}.
\newblock Springer, New York, 2018.

\bibitem[{van der Laan} and Rubin(2006)]{vdl2006targeted}
Mark~J {van der Laan} and Daniel Rubin.
\newblock Targeted maximum likelihood learning.
\newblock \emph{The International Journal of Biostatistics}, 2\penalty0 (1),
  2006.

\bibitem[van~der Laan et~al.(2007)van~der Laan, Polley, and
  Hubbard]{vanderLaan&Polley&Hubbard07}
M.J. van~der Laan, E.~Polley, and A.~Hubbard.
\newblock Super learner.
\newblock \emph{Statistical Applications in Genetics \& Molecular Biology},
  6\penalty0 (25):\penalty0 Article 25, 2007.

\bibitem[van~der Vaart(1998)]{vanderVaart98}
A.~W. van~der Vaart.
\newblock \emph{Asymptotic Statistics}.
\newblock Cambridge University Press, 1998.

\bibitem[{von Mises}(1947)]{mises1947asymptotic}
R~{von Mises}.
\newblock On the asymptotic distribution of differentiable statistical
  functions.
\newblock \emph{The annals of mathematical statistics}, 18\penalty0
  (3):\penalty0 309--348, 1947.

\bibitem[Wager and Walther(2015)]{wager2015adaptive}
Stefan Wager and Guenther Walther.
\newblock Adaptive concentration of regression trees, with application to
  random forests.
\newblock \emph{arXiv preprint arXiv:1503.06388}, 2015.

\bibitem[Young et~al.(2014)Young, Hern{\'a}n, and
  Robins]{young2014identification}
Jessica~G Young, Miguel~A Hern{\'a}n, and James~M Robins.
\newblock Identification, estimation and approximation of risk under
  interventions that depend on the natural value of treatment using
  observational data.
\newblock \emph{Epidemiologic methods}, 3\penalty0 (1):\penalty0 1--19, 2014.

\bibitem[Zheng and van~der Laan(2011)]{zheng2011cross}
Wenjing Zheng and Mark~J van~der Laan.
\newblock Cross-validated targeted minimum-loss-based estimation.
\newblock In \emph{Targeted Learning}, pages 459--474. Springer, 2011.

\end{thebibliography}


\begin{thebibliography}{3}
\providecommand{\natexlab}[1]{#1}
\providecommand{\url}[1]{\texttt{#1}}
\expandafter\ifx\csname urlstyle\endcsname\relax
  \providecommand{\doi}[1]{doi: #1}\else
  \providecommand{\doi}{doi: \begingroup \urlstyle{rm}\Url}\fi

\bibitem[Kennedy(2019)]{kennedy2018nonparametric}
Edward~H Kennedy.
\newblock Nonparametric causal effects based on incremental propensity score
  interventions.
\newblock \emph{Journal of the American Statistical Association}, 114\penalty0
  (526):\penalty0 645--656, 2019.

\bibitem[{van der Laan} and Rose(2011)]{vanderLaanRose11}
Mark~J {van der Laan} and Sherri Rose.
\newblock \emph{Targeted Learning: Causal Inference for Observational and
  Experimental Data}.
\newblock Springer, New York, 2011.

\bibitem[{van der Vaart} and Wellner(1996)]{vanderVaart&Wellner96}
Aad~W {van der Vaart} and Jon~A Wellner.
\newblock \emph{Weak {C}onvergence and {E}mprical {P}rocesses}.
\newblock Springer-Verlag New York, 1996.

\end{thebibliography}
